\documentstyle[11pt,aaspp4]{article}

\lefthead{Klahr et al.}
\righthead{HR 4796A}

\begin{document}

\title{Dust Distribution in Gas Disks.\\A Model for the Ring Around HR 4796A}

\author{H.Hubertus Klahr \and D.N.C. Lin}
\affil{UCO/Lick Observatory, University of California,
    Santa Cruz, CA 95064}



\begin{abstract}
There have been several model analyses of the near and mid IR flux
from the circumstellar ring around HR4796A. In one set of models, the
10 and 18 $\mu m$ IR flux has been attributed to the reprocessing of
stellar radiation by $\mu m$-size particles. Since these particles are
being blown away, on a dynamical time-scale, by the radiation pressure
of HR4796A, they must be continually replenished by the collisional
fragments of larger particles.  If the ring appearance persisted for
the life span ($8 \times 10^6$ yr) of HR4796A, a parent-particle
reservoir with a total mass greater than $300 M_\oplus$ would be needed.  In order
to avoid being conspicuous at longer wavelengths, most of the mass 
must be contained in parent particles larger than $20-40$ cm.  In other 
models, it has been suggested that the IR flux from the rings is emitted 
by sufficiently large particles that survive the radiative blow out 
by their host star.  In a gas free ring, greater than $3.2 \mu m$-size particles
would survive radiative blow out and a total of $10^{-2} M_\oplus$
would be adequate to account for the observed IR flux.  But, in the
vicinity of a young star, the possibility that the dust ring is
embedded within a residual protostellar gas disk cannot be ruled
out. In a gas-rich environment, larger sizes ($>100 \mu m$) are needed
for the particles to survive the radiative blow out.  The total dust
mass required to account for the IR flux is $< 10^{-1} M_\oplus$.  The
combined influence of gas and stellar radiation may also account for
the observed sharp inner boundary and rapidly fading outer boundary of
the ring.  The pressure gradient induced by a small ($10\%$) amplitude
variation in the surface density distribution of a low-mass gaseous
disk would be sufficient to modify the rotation speed of the gas.  The
resulting hydrodynamic drag on modest-size ($>100 \mu$m) particles
would be adequate to compensate for the turbulent stirring, radiative
drag and radiation pressure such that they remain gravitationally bound 
to the system. The required surface density variation of the gas may be
induced by 1) the perturbation of a low-mass planet or the binary
companion HR 4796B, 2) the photo evaporation of the disk, or 3) from the
variations in the viscous angular momentum transport and mass
diffusion rate in the disk.  We show that the structure of the dust ring 
is preserved during and after the gas is being depleted such that similar 
rings may be common among early-type stars.

\end{abstract}




%

\section{Introduction}
The observational discoveries (Smith \& Terrile 1984) of $\beta$ Pic
type dusty rings provide tantalizing hints of residual planetesimal
disks similar to the Kuiper belt may exist around other stars
(Artymowicz 1997).  Perhaps the most intriguing dust rings are those
around HR 4796A (Schneider {\it et al.} 1999) and HD 141569
(Weinberger {\it et al.}  1999).  The sharp surface density variations
near the ring boundaries of HR 4796A are remarkably similar to that
found around Saturn's F ring (Cuzzi {\it et al.} 1984) and Uranus'
$\epsilon$ ring (Elliot \& Nicholson 1984).  Under the action of
viscous stress, these differentially rotating disks have a general
tendency to diffuse in the radial direction (Lynden-Bell \& Pringle
1974).  The expansion of planetary rings are prevented by the tidal
torques of nearby shepherding satellites (Goldreich \& Tremaine 1979,
1982).  The structural resemblance between circumstellar and planetary
rings naturally leads to the conjecture that the former, too, may be
perturbed by nearby hypothetical planets (Jura {\it et al.} 1995).
The reported azimuthal brightness asymmetry (Telesco {\it et al.} 2000)
provides a supporting evidence for the scenario that the ring may be
gravitationally perturbed by a planetary or stellar companion (Wyatt
{\it et al.} 2000).

Another motivation for invoking the embedded planet scenario is the
perseverance of the ring itself.  The NICMOS image shows that the
surface brightness has a maximum at 70 AU with a sharp decline at the
smaller radii and an inner cutoff near 50 AU.  Outside this region,
the surface brightness decreases more gradually with radius to less
than half of the peak brightness at 90AU (Schneider {\it et al.}
1999). The observed ratio of flux from the ring to that from the star
increases with wavelength at 1.1 and 2.6 $\mu m$, indicating the mean
particle size is larger than these wavelength. But, using the flux
ratio at 10 and 18 $\mu m$ and assuming a Mie emissivity, Telesco {\it
et al.}  (2000) inferred a ``characteristic'' diameter for the dust
particles to $\sim 2-3 \mu m$ which is in contrast to the lower
particle-size limit of $10 \mu m$ inferred from the $60 \mu m$
radiation by Jura ({\it et al.} 1993). The observed spectral energy
distribution (SED) with wavelength $\lambda = 10-10^3 \mu m$ (Koerner
{\it et al.} 1998) has been modeled. Using single-size particles ($>100 
\mu m$), Augereau {\it et al.} (1999) were able to fit the
observed SED between $10-10^2 \mu m$ but not the resolved images of
the ring.  Both the SED and the resolved images were better modeled
with reprocessing by particles with a power-law collisional
particle-size distribution (Heller 1970, Mathis {\it et al.} 1977)
with adopted minimum ($a_{\rm min}$) and maximum ($a_{\rm max}$) sizes
of $\sim 10 \mu m$ and a few meters respectively.  In these models,
since the models are applied to SED in the wavelength range of
$10-10^3 \mu m$, the actual values chosen for $a_{\rm min}$ and
$a_{\rm max}$ do not significantly affect the actual match.

With the stellar luminosity $L_\star = 35 L_\odot$, small particles,
if they exist, would be blown away by the radiation pressure on the
time-scale of a few hundred years (Backman \& Paresce 1993).  In the
models adopted by Augereau {\it et al.} (1999), all particles with
sizes larger than $a_{\rm min}$ would survive the radiative blow-out
effect.  But for the models proposed by Telesco {\it et al.} (2000),
the ``characteristic-size'' particles would not be able to remain.  In
a companion paper, Wyatt {\it et al.} (2000) suggested that these
small particles are probably the collisional fragment of a population
of larger particles.  In this replenishment scenario, it is still
essential for the parent particles to have an adequate reservoir of
total mass to replenish the $\mu m$-size particles for HR 4796A's life
span $\tau_\star = 8 \times 10^6$ yr (Stauffer {\it et al.} 1995,
Jayawardhana {\it et al.} 1998).  In order to be consistent with the
observed SED and ring structure, most of these parent particles must
also be located within $70 \pm 20$ AU and have a surface density
comparable to or less than that inferred by Augereau {\it et al.}
(1999). One or more nearby planets may provide both the supply and
confinement mechanism for the parent particles.

An important issue to be resolved is the required mass for such
perturbers.  In HR 4796A, such confining planets must either form or
migrate to the vicinity of the rings (which is well beyond the orbit
of Neptune) within $\tau_\star$.  In conventional theories, the first
stage of planetary formation is the emergence of planetesimals and
solid cores (Safronov 1969; Wetherill 1980).  Only after these cores
have acquired a few Earth-masses does dynamical accretion of gas become
possible (Pollack {\it et al.} 1996).  But, beyond the orbit of
Neptune, the formation of a few Earth masses core requires at least
several $10^7$ yr ($> \tau_\star$) unless the mass of solid material
is much larger than that inferred from the minimum mass solar nebula
model (Lissauer 1993).  Assuming an initial surface density 10-20
times that of the minimum mass solar nebula (which corresponds to
150-300 $M_\oplus$ within the annulus between $\sim 58-82$ AU), Kenyon
{\it et al.}  (1999) found it possible to form planetesimals larger
than $10^3$ km within $10^7$ yr. They further suggested that
gravitational perturbation by these large objects may excite large
velocity dispersion that may lead to disruptive collisions.  However, if
this assumed surface density of the dust particles is augmented for
the solar composition, the inferred mass of the gaseous ring ($\sim
0.05-0.1 M_\odot$ within $70 \pm 12$ AU) would make it gravitationally
unstable.  Cameron (1978) and Boss (1998) have suggested that
gravitational instability in the disk may indeed lead to the formation
of giant planets directly from the disk gas.  But, some numerical
simulations show that the growth of non axisymmetric perturbations
excited by gravitational instability are more likely to induce
efficient transfer of angular momentum in the disk rather than
fragmentation (Nelson {\it et al.}  1998, Burkert \& Bodenheimer 1996).
The initial surface density required for this conjecture is also large
compared with the mass inferred for the outer regions of typical
protostellar disks (Beckwith \& Sargent 1993a).

In this paper, we examine the persistence of the dust ring structure
around HR 4796A.  Since the host star is only $8 \times 10^6$ yrs old,
we hypothesize the existence of some residual gas analogous to those
found around several young stellar objects with similar ages (Beckwith
{\it et al.} 1990, Beckwith \& Sargent 1993b, Zuckerman {\it et al.} 1995,
Hartmann {\it et al.} 1998).  An upper limit 1-7 $M_\oplus$ of gas has been
inferred from the mm continuum radiation by Greaves {\it et al.}
(2000).  (A larger gas mass may exist if a substantial amount of heavy
element is depleted from the gas phase and is stored in particles
larger than a few mm.)  We explore the interaction between the dust,
radiation, and residual gas in the disk.  In \S2, we briefly
recapitulate the physical interaction between particles, stellar
radiation, and residual gas.  We show that the hydrodynamic drag tends to
promote the effect of radiation pressure and increase the critical
particle size for radiative blow out. In \S3, we provide some
constraints on the size and total mass of particles in the ring around
HR 4796A. We show that variations in the surface distribution of the
gas provide a favorable condition for modest (more than 100 $\mu m$)
size particles to congregate into a dust ring. This ring structure is
preserved during and after the depletion of the disk gas.  We suggest
that these particles provide a supply of the dominant contributors to
the IR emission of the dust ring.  Finally in \S4, we discuss the
implications of these results on planet formation and the ubiquity of
similar rings around other early type stars.

\section{The Dynamics of Residual Dust Particles}
In this section, we examine the interaction between stellar radiation,
residual disk gas, and dust particles of various sizes.  

\subsection{The gaseous disk}
For illustrative purpose, we consider a simple model in which a disk
of dust particles is embedded in a residual gaseous disk. The only
free parameters in this model are those which describe the gas surface
density $\Sigma(R)$ as a function of the disk radius $R$. All other
gas properties are deduced in a self-consistent way.

The qualitative results presented here depend on the gross global gas
distribution in the disk and are insensitive to the detailed expression of
$\Sigma (R)$.  In principle, the $\Sigma$ distribution is determined
by the efficiency of angular momentum transfer (e.g.\ via viscosity from
turbulence) in the disk and the
tidal perturbation by planetary or stellar companions.  The former
mechanism generally leads $\Sigma$ to attain a power law function of
$R$ in most regions of the disk (Lin \& Bodenheimer 1981; Ruden \&
Lin 1986). The latter process leads to a localized $\Sigma$ variation
such as gaps (Lin \& Papaloizou 1979; 1993) and confined rings
(Goldreich \& Tremaine 1978).  Although low-mass companions may not be
able to induce the formation of a gap, they can nevertheless introduce
variations in the surface density of the disk.
Self-gravitational effects will not affect the disk structure
as the surface density in our models is always orders of
magnitude smaller than any critical value (see Toomre 1964).
For computational
purpose (e.g.\ to produce our plots), we adopt the following generic prescription in which the
global variation of $\Sigma$ is non monotonic:
\begin{eqnarray}
\Sigma \: & = &\Sigma_0 \times
\: \left(
\begin{array}{ccc}
d \left(\frac{R}{R_0}\right)^{-2.5}
& if: & R < R_1\\
e^{-\frac{(R-R_0)^2}{\delta R_0^2}}
& if: & R_1 \leq R \leq R_0\\
\left(\frac{R}{R_0}\right)^{-2.5}& if: & R > R_0
\end{array} \; \; \right)\;
\end{eqnarray}
where $d$ is a model parameter which determines the 
amplitude of a local maximum. 
The magnitude of $\Sigma$ at some fiducial radial
location $R_0$ is set to be $\Sigma_0$. For $R < R_1$ and $R > R_0$,
$\Sigma$ decreases monotonically with $R$ following a power law (with
an index -2.5) which is somewhat steeper than that for the minimum
mass solar nebula model (Hayashi {\it et al.} 1985).  But in the outer
region of the disk, $\Sigma$ is likely to decline more rapidly with
$R$ as inferred from observation (Mundy {\it et al.} 1996, Hartmann {\it et al.} 
1998). Since our results are insensitive to the detailed nature of the global
radial dependence of $\Sigma$, we choose such a power law so that the
disk mass remains finite at large $R$.

The inner and outer regions of the disk are connected through a
transition zone at $R_1 \leq R \leq R_0$.  
In this zone the surface density has the shape of a
Gaussian with a width of $\delta R_0$.
The value of $R_1$ is set
to satisfy $(R_1 - R_0) = \delta R_0 (2.5 {\rm ln} ( R_1/R_0) - {\rm
ln} d)^{1/2}$ such that $\Sigma$ is continuous. In all our models, we
choose $d < 1$ which is reasonable because the continuum flux in the
mid-IR wavelength range is observed to be more intense than that in
the near-IR range in several protostellar disks (Hillenbrand {\it et
al.} 1998).  One possible interpretation is that these disks are more
depleted in the inner region or better preserved in the outer region
as in the context of circumbinary disks (Jensen \& Mathieu 1997).  On 
the theoretical side, such a $\Sigma$ variation is expected if a) the 
viscosity decreases rapidly with radius (Ruden \& Pollack 1991), b) 
the inner region is depleted by either a stellar or a disk wind (Shu 
{\it et al.} 1994, Konigl \& Ruden 1993), c) HR4796A is a A0V
star and a strong source of UV radiation which could lead to photo
evaporation in some regions of the disk (Shu {\it et al.} 1993) and
quench gas infall along the rotation axis and onto the inner regions
of the disk (Yorke {\it et al.} 1995), d) the disk is depleted by the
formation of one or more planets which are expected to first emerge in
the inner regions of the disk (Jura {\it et al.} 1993), or e) the
structure of disk around HR4796A may be tidally perturbed by the
stellar companion HR4796B (Augereau {\it et al.}  1999), especially if
the binary orbit is eccentric (Korycansky \& Papaloizou 1995).  The
difference between the inner and outer regions of the disk can be
minimized by setting $d$ close to unity.  Finally, the $\Sigma$
variation in the transition region $R_1 \leq R \leq R_0$ can also
represent a tidal perturbation of a nearby embedded planetary
companion.

The value of $\Sigma_0$ is set such that the mass $M_g = 2 \pi
\int_{R_0 - \delta R_0} ^{R_0 + \delta R_0} R\Sigma(R)dR$ contained
within a ring centered on $R_0$ with a half width $\delta R_0$ is to
be specified as a model parameter. This mass is to be spread over a
ring with a half width $\delta R_0 = 10-20$ AU centered on a radial
location $R_0 = 70$AU.  In our models $M_g$ is 1-100 Earth
masses ($M_\oplus$). The corresponding $\Sigma$ is orders of magnitude
smaller than that for the minimum mass solar nebula model at a similar
distance from the central star (Hayashi, Nakazawa \& Nakagawa 1985).

The outer regions of the disk are optically thin such that every dust
particle in the disk receives direct irradiation from the central
object. The particles are heated to the local black body temperature
(small grains may be hotter because they cannot radiate efficiently,
Chiang \& Goldreich 1997).  The luminosity and mass of HR 4796A are
$L_{\star} = 35 L_{\sun}$ and $M_{\star} = 2.5 M_\odot$ respectively
(Jura {\it et al.} 1993). The temperature of the dust grains can be 
given by:
\begin{equation} 
T_d = \left(\frac{L_{\star}}{16 \pi a R^2}\right)^{1/4}
\end{equation}
where $a$ is the radiation density constant. 
This upper limit estimation holds for particles
larger than the typical wave length of the radiation
and assuming an albedo of zero.  A more careful treatment
of the optical properties of the particles is not necessary
in the context of the effects studied in this paper.
 
If gas molecules collide frequently with the dust particles, they can
also attain a similar temperature, $T_{high} = T_d$.  In the low gas
density limit, however, thermal conduction between the dust and the
gas is less efficient than the gas radiative processes.  As a
consequence of this thermal imbalance, the gas temperature may
decrease to a minimum value ($T_{low}=10$K) which is comparable to
that of the radiation background field (Preibisch, Sonnhalter and
Yorke 1995; Yorke \& Lin in preparation).  In general,
$T_{low}<T_{gas}<T_{high}$ and the associated sound speed $c_s =
\sqrt{T_{gas} * R_{gas} / \mu}$, where $R_{gas}$ is the gas constant
and $\mu$ the mean molecular weight (i.e. $\mu = 2.353$ for an
$He-H_2$ mixture). The disk thickness is determined by its pressure
scale height $H_p = c_s / \Omega$ where $\Omega \equiv (G M_\ast /
R^3) ^{1/2}$ is the Keplerian frequency.  At $R=70$ AU, $0.03 < H_p/R
< 0.1$ for the low and high temperature limits.  For the parameters we
have adopted, the disk gas is likely to be in the low temperature and
small thickness range.  The local gas density in the midplane of the
disk can be estimated as $\rho = \Sigma/2 H_p$ and the pressure $p$
becomes $p = c_s^2 \rho$.

The azimuthal orbital frequency of the gas ($\omega$) is given by the
condition for hydrostatic equilibrium between the gravitational
potential of the central object and the sum of centrifugal
acceleration and the radial pressure gradient
\begin{equation}
\frac{G M_\star}{R^{2}} = \omega^2  R - \frac{1}{\rho}\nabla{p},
\label{eq:ome}
\end{equation}
which leads, in first order approximation, to a local deviation $dV$
from the Keplerian profile of (see Whipple 1972)
\begin{equation}
dV \equiv (\omega-\Omega) R  = V_\phi - \Omega R = \frac{1}{2 \Omega 
\rho}\nabla{p} \simeq \frac{c_s ^2}{2 \Omega R} \frac{\partial {\rm ln}
\rho}{\partial {\rm ln} R} 
\end{equation}
where $V_\phi = \omega R$ is the azimuthal speed.  In \S2.4, we 
consider the case that this angular frequency may be modified by the
particles' drag when they become the dominant constituent of the disk. 

Results under specific assumptions for the free parameters of 
this model will be discussed in the \S 3.

\subsection{Radial drift of dust grains}
The particles are drifting radially under two independent physical
effects: 1) hydrodynamic drag by the gas and 2) the radiation pressure
of photons emitted by the central object.

\subsubsection{Departure from Keplerian flow}
While the pressure gradient induces the gas to attain a finite $dV$,
the particles' motion is determined by the gravity of the central
star, the hydrodynamic drag by the disk gas, and the drag and pressure
by the stellar radiation.  Radial inward and outward drift of the
particles can be generated by the non-Keplerian motion of the
gas. Small particles are well coupled to the gas, thus they also move
on non-Keplerian orbits provided their Stokes number $S\!t \equiv
\tau_f \Omega << 1.0$ (Weidenschilling 1977).  The friction time for
a spherical particle in the Epstein regime (e.g.\ Klahr \& Henning
1997) is given by
\begin{equation}
\tau_f = \frac{a_d \rho_d}{c_s \rho} = \frac{3 m_d}{4 A_d c_s \rho},
\end{equation}
where $a_d$ is the particle's radius, $m_d$ its mass, $A_d= \pi a_d^2$
its cross section, $\rho_d$ its density, $c_s$ the thermal velocity of
the gas and $\rho$ the local gas density.  In the case of transonic
velocities (which is irrelevant in our model) $c_s$ has to be replaced
by $\sqrt{c_s^2 + dv^2}$ with the relative velocity ($dv $) between
the particle and the gas (H.\ Yorke 1999, private communication).  The critical particle size
which demarcates the small and large particle size range then becomes
\begin{equation}
a_0 = \Sigma / 2 \rho_d
\label{a_0}
\end{equation} 
which is a few cm for a minimum mass solar nebula model at 70 AU and
smaller for a depleted gaseous disk.  In our standard models 
(see Sect.\ \ref{Sect_modA.ref} and Sect.\ \ref{Sect_modB.ref} ), $a_0 =
0.06$ cm at $R_0$.

In the solar system today, the absence of gas makes $St >>1$ for all
particles.  Modest-size particles spiral inwards due to the Poynting
Robertson effect (see below), but spherical particles (with $\rho_d = 2.5$gm
cm$^{-3}$) with size
\begin{equation}
a_d < a_1 \equiv \frac{3}{16 \pi} \frac{L_\star}{G M_\star c \rho_d} \sim 
3.2 \mu m
\label{eq:a11}
\end{equation}
would be blown out by the radiation pressure on hyperbolic orbits as
``beta meteorites'' (see Gr\"un {\it et al.} 1985) on a time-scale comparable
to the local dynamical time-scale $\tau_d \sim \Omega^{-1} \sim 10^2$
yr (see below). 
For our considerations we neglect particles smaller than the
typical wavelength of the radiation (No Mie scattering).

 Larger particles would remain bound to the host star.
They can even remain on a circular orbit with a sub Keplerian
azimuthal velocity
\begin{equation}
v^*_\phi = \sqrt{\frac{G M_\star-\frac{L_\star A_d}{4 \pi m_d c}}{R}},
\label{eq:nearKep}
\end{equation}
where $v^*_\phi$ denotes the particles theoretical velocity in the
absence of gas, which is an important quantity (see below).
While the particles are orbiting around the central star they also
experience a radiative head wind due to the aberration of its photons.
In gas-free environments, such as our own solar system, this
Poynting-Robertson radiative drag effect induces the azimuthal
velocity ($v_\phi$) of small particles to change at a rate
\begin{equation}
\dot v_\phi = - \frac{L_{\star}}{4 \pi R^2} \frac{A_d}{m_d c^2} v_\phi .
\end{equation}
This process leads an inward drift for all remaining particles with 
$a_d > a_1$.

In a gaseous environment, the combined influence of hydrodynamic and
radiative drag induces the particles' azimuthal velocity to change at
a rate
\begin{equation}
\dot v_\phi = -\frac{(v_\phi-V_\phi)}{\tau_f} - \frac{L_{\star}}{4 \pi R^2} 
\frac{A_d}{m_d c^2} v_\phi. 
\end{equation}
If $\tau_f$ is small in comparison to the relevant evolutionary
time-scales (see Fig.\ \ref{modA.ref} and Fig.\ \ref{modB.ref}), 
an equilibrium would be rapidly established
with $\dot v_\phi = 0$ so that
\begin{equation}
v_\phi =  \frac{V_\phi}{1 +\tau_f  \left(\frac{L_{\star}}{4 \pi R^2} 
\frac{A_d}{m_d c^2}\right)}.
\end{equation}
Substituting $\tau_f$, we find
\begin{equation}
v_\phi =  \frac{V_\phi}{1 + \left(\frac{L_{\star}}{4 \pi R^2} 
\frac{3}{4 c^2 c_s \rho}\right)}
\label{eq:vphi}
\end{equation}
which is independent of particle sizes (as long as $S\!t << 1$ and
the particles are larger than the wavelength of the radiation)!

Around young stellar objects where the particles are embedded in a
gaseous circumstellar disk environment, the term $dv \equiv v_\phi -
V_\phi$ (see Fig.\ \ref{modA.ref}) is typically less than of 1 mm/s at 70 AU while
$dV \sim 10^{3}$cm/s. Note that the Poynting Robertson effect is
insignificant until $\Sigma$ is depleted below $3 L_\star / 8 \pi R^2
c^2 \Omega$ which is not only more than four orders of magnitude below
that for our standard model and that inferred from mm continuum
radiation for typical protostellar disks but also smaller than that
inferred for the CO gas around some young stellar objects (Zuckerman
{\it et al.} 1995). (The actual gas content may be even larger if a significant
amount of CO gas is depleted due to grain condensation.)  Thus, in our
estimate of the particles' motion, we only consider contributions from
$dV$ and neglect the Poynting Robertson effect.

\subsubsection{Radial drift}
In addition to the radiative drag in the azimuthal direction, the
stellar photons also exert on the particles a radiative pressure in
the radial direction.  In general, the radial drift $v_r$ of small
particles (with $St <1$) results from the non-Keplerian motion and the
radial radiation pressure can be obtained from the radial component of
the equation of motion such that
\begin{equation}
\frac{\partial v_r}{\partial t} = -\frac{v_r}{\tau_f} + \frac{{v_\phi}^2}{R} 
- \frac{G M_\star}{R^2} + \frac{L_{\star}}{4 \pi R^2} 
\frac{A_d}{m_d c}.
\end{equation}
In order to simplify work we define $\Omega^* (\equiv
v_\phi ^*/R)$ which is the orbital frequency with respect to the
radial radiation pressure (see Eq.\ \ref{eq:nearKep}).  Note that
$\Omega^* < \Omega$ in general.  Thus it follows
\begin{equation}
\frac{\partial v_r}{\partial t} 
= -\frac{v_r}{\tau_f} + \frac{\left(\Omega^* R +
dV^*\right)^2} {R} - {\Omega^*}^2 R.
\end{equation}
Here $dV^* \equiv v_\phi - \Omega^* R$ denotes the deviation of the
gas from the modified Keplerian orbit.  Since the radius dependence of
gravity and radiation pressure is identical, the outward radiation
pressure effectively reduces the gravitational attraction on the
particles imposed by the host star everywhere (Gustafson 1994).

The ``characteristic size'' of particles inferred by Telesco {\it et
al.} (2000) is smaller than $a_1$ such that they are expected to be
blown away by the radiation pressure of HR4796A.
In order to remove this dilemma, Wyatt {\it et al.}
(1999) suggested that although these small particles are continually
being blown out, they are also being replenished by collisional
fragmentation of larger particles which are not significantly affected
by the radiation pressure.  In a gas free environment, the parent body
size must be larger than $a_1$.

But, around many young stellar objects, small particles are
embedded in a residual gaseous background. The gas generally does not
directly experience the radiation pressure because its molecular
opacity is too small.  Thus, the gas azimuthal speed ($V_\phi$) is
independent of $L_\star$ and it is given by eq.\ (\ref{eq:ome}).  The
discussions in \S2.2.1 indicate that particles with $St < 1$ (or
equivalently with $a_d < a_0$) have $v_\phi \sim V_\phi$ (see
eq. {\ref{eq:vphi}}).  As the hydrodynamic drag forces these small
particles to nearly corotate with the gas, they attain a finite $v_r$
as a consequence of being out of hydrostatic equilibrium.  When a
steady state is established,
\begin{equation}
v_r = \tau_f 2 \Omega^* dV^* \,\,\,\,\,\,{\rm for}\,\,\,\,\,\, St << 1,
\label{Eq_v_r}
\end{equation}
in the limit that the quadratic term $dv^2$ is small in comparison to
$\Omega^* R dv^*$ (cf Weidenschilling 1977; He used another definition
of $dV$!). The sign of $dV^*$ can now be negative or positive,
depending on whether the particle is less or even more sub-Keplerian than 
the gas.  Relatively large particles follow in a similar fashion,
\begin{equation}
v_r = dV^* \,\,\,\,\,\,{\rm for}\,\,\,\,\,\, St = 1,
\label{Eq_v_r1}
\end{equation}
and
\begin{equation}
v_r =  \frac{2  dV^*}{\Omega^*\tau_f}\,\,\,\,\,\,{\rm for}\,\,\,\,\,\, St >> 1.
\label{Eq_v_r>}
\end{equation}

Note that the effect of radiation pressure decreases with the
particle size.  From this equation one can deduce that in regions
where $dV >0$ (as a positive of a negative gas pressure gradient), all the
particles would migrate outwards since their
azimuthal speed is always slightly hyper-Keplerian, just like the gas.  
But in the more common situation where $dV \leq 0$ (which corresponds to a
negative gas pressure gradient as expected in unperturbed regions of
protostellar disks), hydrodynamic drag is more important than
radiative drag for all particles with
\begin{equation}
a_d > a_2 \equiv -\frac{3 L_{\star}}{32 \pi R^2 c \Omega \rho_d dV}
\label{eq:a_2}
\end{equation}
so that they undergo orbital decay. In contrast, particles with $a_d
< a_2$ are blown away by the stellar radiation pressure (see Fig.\ \ref{modA.ref} and Fig.\ \ref{modB.ref}). 
This minimal size for the given parameters of HR 4796A is $ a_2
\sim 100-200 \mu$m for almost all disk radii $R$. Only at those
location where variations in $\Sigma$ lead to a reduction in $dV$ is
$a_2$ greater than $500 \mu$m.  In the location where $dV $ vanishes or attains
a negative value, $a_2$ becomes infinite and particles of all sizes
(including planets) are prevented from inward radial drift.

We note that the magnitude of $a_2$ is larger than $a_1$ in
eq(\ref{eq:a11}) by a factor $\sim (R^2/H_p^2) (d {\rm ln} \rho / d
{\rm ln} R)^{-1} \sim 10-100$ because the small particles are forced
to corotate with the gas such that a centrifugal balance is always
maintained.  Thus, if they are embedded in a residual gaseous disk,
the ``characteristic-size'' ($\mu m$-size) particles in HR 4796A
inferred by Telesco {\it et al.} (2000) must be supplied by parent
bodies which are at least two order of magnitude larger than them.

In a gaseous disk environment, all particles with $a_d$ smaller than
both $a_0$ and $a_2$ are blown away by the radiation pressure on the
dynamical time-scale $\tau_d$.  In the limit that $a_0 > a_2$,
particles with sizes $a_0 > a_d > a_2$ can achieve both positive and
negative radial drift velocities depending on the local gas pressure
gradient.  In \S3, we present numerical results to show that these
changes in the direction of particles' radial drift lead to the
particles' accumulation into rings.  Note that the value of $a_2$ does
not directly depend on $\Sigma$.  As the disk gas is depleted, neither
$d V$ nor $a_2$ are strongly affected.  But, $a_0$ is linearly
proportional to $\Sigma$.  When $a_0$ decreases below $a_d$ of some
particles, they decouple from the gas and their azimuthal
velocity becomes closer to the local modified Keplerian value
$v_\phi^*$ (see eq.\ \ref{eq:nearKep}). Since their $a_d$ is larger
than $a_2$ initially and $a_2$ decreases, the
radiation pressure and the Poynting Robertson effect have a weaker
effect than the hydrodynamic drag throughout the evolution such that
their radial drift velocity is described by eq.\ \ref{Eq_v_r>}.
Eventually, the depletion of the residual gas increases $\tau_f$ and
reduces $v_r$.

\subsection{Constrains on the disk profile}
We already noted that the trapping mechanism for particles is located
at places, where the disk becomes Keplerian, as the radial pressure
gradient vanishes, {\it i.e.}
\begin{equation}
\nabla p = \nabla \left(c_s^2 \rho \right) = \nabla \left(c_s \Sigma \Omega \right) = 0.
\end{equation}
Since $\Omega \sim R^{-3/2}$, the general criterion for the ring formation is
\begin{equation}
c_s \Sigma \propto R^{3/2}.
\end{equation}
This criterion can now be used to determine the local shape
of the Surface density profile. We consider three distinct cases: 1)
an isothermal disk($T=T_{low}$), 2) an optically thin disk in radiation
equilibrium ($T=T_{high}$), and 3) an optically thick viscous disk.

In the isothermal case $c_s$ is constant such that the trapping
mechanism requires $\Sigma$ to increase locally as $R^{3/2}$
regardless its magnitude and radial extent.  In the standard optical
thin case, (perhaps most relevant to the case of HR4796A), $c_s \sim
R^{-\frac{1}{4}}$ as $T \sim R^{-\frac{1}{2}}$.  The necessary
power-law $R-$dependence in $\Sigma$ is only slightly steeper than in
the previous case: {\it i.e.} $\Sigma \sim R^{\frac{7}{4}}$.

In an optically thick disk which is heated by the local viscous
dissipation, the disk's mid plane temperature $T_c =
T_e\left(3\tau_r/8 \right)^{1/4}$ where the vertical optical thickness
$\tau_r \approx \kappa \Sigma/2$ and the Rosseland mean opacity
$\kappa = \kappa_o T_c^2$ (Lin \& Papaloizou 1985).  If the disk is
approximately in a steady state, its effective temperature $T_e
\propto R^{-\frac{3}{4}}$ (e.g.\ Lynden-Bell \& Pringle 1974, Bell
{\it et al.} 1997) such that $T_c \propto \kappa_o ^{1/2} T_e ^2
\Sigma^{1/2} \propto R^{-3/2} \Sigma^{1/2}$.  Particles are trapped in
regions where $\Sigma \propto \kappa_o ^{-1/5} R^{9/5}$ which is
similar to the isothermal and the optically thin cases.

The dominant contributors to the disk opacity are grains.  Across some
critical locations (dust destruction zone) in the disk where the
grains sublimate, the opacity increases mainly through changes in the
scaling constant $\kappa_o$ (Ruden \& Lin 1986).  Particles may be
trapped in regions where $\kappa_o \propto R^9$.  A full 2-D detailed
model is needed to address the issue whether such a transition may
indeed be a viable effect to prevent dust in form of meter sized solid
bodies from drifting into the sun and help collect bodies at certain
transition radii from the sun to form the terrestrial planets.  For the 
present context, the ring around HR 4796 A is well outside the ice
and silicate destruction zones.

\subsection{Particle-dominant flow}
In the above discussion, we neglect the feedback effect of
hydrodynamic drag on the gas flow.  In principle, the gas flow may be
perturbed by the motion of the particles when the particles dominate
in mass surface density ($\Sigma_d$) over the surrounding gas
(Nakagawa, Sekiya, \& Hayashi 1986). In the Kuiper Belt today, no sign of residual gas has
been detected.  Gas may be preferentially depleted prior to the
depletion of the dust particles.  When the surface density of the gas
($\Sigma$) decreases below that of the dust ($\Sigma_d$), the hydrodynamic
drag between them may induce the gas to comove with the particles rather
than forcing the particles to corotate with the gas.  In reality, however,
the only particles which can provide an efficient momentum transfer
with the gas are those with sizes $a_d << a_0$ (Nakagawa{\it et al.} 1986).
During the depletion of gas, $a_0$ decreases with $\Sigma$, leading to
a reduction in the total fraction of particles which are well coupled
with the gas.  Thus, the gas speed is unlikely to be significantly
affected by the particles' drag effect.

For those regions of the disk with $\Sigma > \Sigma_d$, gas flow would
be affected by the hydrodynamic drag of the particles near the
midplane of the disk if the total mass density of the particle $n_d
m_d$ exceeds the gas density $\rho \sim \Sigma / H_p$ (Weidenschilling
\& Cuzzi 1993).  (Similarly, particle accumulation due to variations
in the radial drift speed may also lead to local enhancement of
particle density.)  The average mass density of the particles
$\rho_{\rm dust}$ is determined by their scale height $H_d
=\sigma_d/\Omega$ where $\sigma_d$ is the velocity dispersion of the
particles. In a turbulent medium, the particles are being stirred by
the dispersive motion of the gas.  Since the turbulent speed of the
gas is unlikely to exceed its sound speed, particles with $a_d > a_0$
would sediment toward the mid plane and attain a scale height $H_d <
H$.  With a substantially sub sonic turbulent speed, sedimentation is
also possible for sufficiently small ($a_d < a_0$) particles which
couple well with the gas. Only after $\rho_{\rm dust}$ of the small
($a_d < a_0$) particles has exceeded the gas density $\sim \Sigma/2
H$, would the gas motion be forced to corotate with the small
particles' $\Omega^*$ of the rather than $\omega$.  Without a
significant differential motion between them and the gas, particles
with $a_1 < a_d < a_0$ would not undergo any further orbital
migration.  But larger ($a_d > a_0$) particles are less affected by
the radiation pressure and have a larger $v_\phi^*$.  Gas drag induced
by the residual headwind would continue to drive the larger particles
inward though at a much reduced pace.

But, the optical depth of HR4796 A is well below unity and 
\begin{equation}
\rho_{dust} = \frac{2 L_{disk}}{3 \pi L_{*}} \frac{\rho_d a_d}{H_d}
\end{equation} 
(see eq.\ \ref{eq:parden} below).  For particles of $a_d = 200\mu$m the
density is $ \rho_{dust} << 6.3 \times 10^{-17}$g cm$^{-3}$, which is well 
below the gas
density inferred from the minimum mass solar nebula model.  For those
models we have adopted in this paper, particles have negligible 
drag effect on the motion of the gas.

\section{Models for HR 4796A}
In this section, we present a scenario for the ring structure around HR
4796A.  First, we revisit the two existing models for particle size 
distribution by analyzing the preferred size of particles for retention. 
We describe a simple quasi static model in which we assume
that HR 4796A had initially a circumstellar gas and dust disk around
it. We then show that a dust disk will evolve into a ring structure
with sharp edges.  We argue such a structure would be preserved after
the gas is depleted.

\subsection{Replenishment of a population of $\mu m$-size particles}
First, we examine the implication of the scenario proposed by Telesco
{\it et al.} (2000) in which the dominant contributors to the 10 and
18 $\mu m$ radiation from the ring around HR 4796A are $\mu m$-size
particles.  Since the ``characteristic'' size of the particle is
comparable to the wavelength of the photon, the reprocessed radiation
by the disk particles
\begin{equation}
L_{disk} = Q
L_{\star} \frac{A_d}{4 \pi R^2} 
\frac{M_{char} }{m_d}
\end{equation}
where $Q$ is the albedo and assumed to be one for large particles
and $dR$ the radial width of the dust ring.
Assuming $SiO$-dust particles which have spherical shape and a density
$\rho_d = 2.5 $gm cm$^{-3}$, the total mass of those particles with
``characteristic'' sizes is
\begin{equation}
M_{char} = \frac{16 \pi}{3} \frac{L_{disk}}{L_{\star}} R^2 
\rho_d a_d.
\label{eq:mchar}
\end{equation}
The total IR luminosity of the ring is estimated to be
$L_{disk} / L_\star = 5 \times 10^{-3}$ (Jura {\it et al.} 1993) so that
\begin{equation}
M_{char} \approx  (a_d / 1 {\rm cm}) \times 2.3 \times 10^{29} g \approx 
(a_d / 1 {\rm cm}) \times 38 M_\oplus.
\label{mpart}
\end{equation}
For the all particle size to be at the ``characteristic'' size 
$a_c = 1-2 \mu m$, one would need $M_{char}\sim 4-8 \times 10^{-3} M_\oplus$.

If the IR emitting particles are smaller than both $a_1$ and $a_2$
they would be blown away from the ring region on a dynamical time-scale
$\tau_d = \Omega^{-1} \sim 10^2$ yr in both gas-rich and gas-free
environment.  Over the life span of HR4796A ($\tau_\star = 8 \times
10^6$ yr), an amount $M_{lost} ^{char} \approx M_{char} \Omega \tau_\star
\approx 300 M_\oplus$ of characteristic-size particles is lost through
blow out by radiation pressure.  In order to replenish this loss,
Wyatt {\it et al.} (2000) proposed that the IR-emitting $\mu m$-size
particles are collisional fragments of larger parent particles.
Unless we are observing HR 4796A in a special epoch, the total amount
of mass in the parent particles ($M_{tot}$) is likely to be comparable
to or larger than $M_{lost} ^{char}$.

In the model presented by Telesco {\it et al.} (2000), only a single
population of characteristic $\mu m$-size particles is assumed to be
responsible for the reprocessing of stellar radiation.  In order for
the parent particles assumed by Wyatt {\it et al.} (2000) to remain
inconspicuous, their total surface area must be less than that of the
$\mu m$-size IR-emitting particles.  For a population of parent 
particles to contribute a major fraction of $M_{tot}$, their total 
area is $\propto a_d^{-1}$.  Since $M_{lost} ^{char}/ M_{char} \sim 
\Omega \tau_\star \sim 10^5$, the minimum size for the parent 
particles is $a_{min} \sim 10^5 a_c \sim 10 cm$ such that their 
$\tau_z$ is less than that of the IR-emitting $\mu m$-size particles. 

\subsection{Particle collisions}
The production of $\mu m$-size particles requires collisions.
The collisional frequency for individual characteristic particles is
\begin{equation}
\nu_{c} \sim n_d A_d \sigma_d,
\label{eq:omec}
\end{equation}
where $\sigma_d$ is the velocity dispersion of the particles.
The spatial number density of those particles which contribute to most
of the IR radiation is
\begin{equation}
n_d = \frac{M_{char}}{m_d 4 \pi R dR H_d} \simeq \frac{4 R}{3 dR}
\frac{L_{disk}}{L_{*}} \frac{\rho_d a_d}{m_d H_d}.  
\label{eq:parden}
\end{equation}
For photons with wavelength smaller than the characteristic size of
the particles, the particles' opacity is determined by their geometric
cross section $A_d$.  The optical depth in the direction normal to the plane
of the disk is $\tau_z \sim (M_{char}/m_d) (A_d /2 \pi R dR) \sim
L_{disk}/L_\star$.  From Eqs. (\ref{eq:mchar}) and (\ref{eq:omec}) we find 
the collisional frequency for IR emitting particles to be
\begin{equation}
\nu_{c} \sim \nu_{cI} = \frac{R}{dR} 
\frac{L_{disk}}{L_\star} \Omega 
\sim \tau_z \Omega < < \tau_d^{-1}.
\label{eq:omegac}
\end{equation}
If the dominant contributors of the IR radiation are $\mu m$-size
particles as postulated by Wyatt {\it et al.} (2000), they would be
blown out on a dynamical time-scale because their $a_d < a_1$ (in a
gas-free environment) and $a_d < a_2$ (when the particles are embedded
in a gaseous disk).  Thus, after they are produced from the collisions
of their parent particles, these particles would not have sufficient
time to collide, fragment, or coagulate among themselves before they
are ejected of the ring region. Furthermore, in order for the
$\mu m$-size particles to remain as the dominant IR emitters, both
$\tau_z$ and $\nu_{c}$ must decrease with particle size so that the
particles with $1-2 \mu m < a_d <a_1$ or $a_2$ would not dominate
the radiation reprocessing.  Since both $a_1$ and $a_2$ are considerably
larger than a few $\mu m$, the hypothetical $\mu m$-size IR-emitting
particles must be produced directly (rather than through a cascade
process) during collisions between the smallest surviving particles
(those with $a_d > a_1$ or $a_d > a_2$).

We now consider the collision frequency of the population of parent
particles which are the main suppliers of the $\mu m$-size IR-emitting
particles.  In the minimum-size (with $a_d \sim a_{min} \sim 10$ cm)
limit, the parent particles' $\nu_{c}$ is comparable to that of the
IR-emitting particles, {\it i.e.} $\nu_{cI}$ in
eq(\ref{eq:omegac}). On the dynamical time-scale during which the $\mu
m$-size particles are blown away, a total fraction $\nu_{cI}
\tau_d$ of the parent particle population would collide.  In order for
this limited number of events to directly replenish the mass loss due
to the radiative blow-away of the IR-emitting particles, each
collision must generate $N_f \sim (\tau_\star/\tau_d)^2 L_\star/L_{disk} 
\sim 10^{12}$ ($\mu m$-size) fragments. The parent bodies could have
a size $a_d > a_{min}$, in which case, collisions would occur less
frequently by a factor $\sim a_{min}/a_d$ provided they can lead to
the production of $\sim (\tau_\star/\tau_d)^2 (a_d L_\star /a_{min}
L_{disk})$ fragments. Note that particles with sizes $> (\tau_\star
L_{disk} / \tau_d L_\star) a_{min}$ collide less than once during
$\tau_\star$.  Nevertheless, at least in principle, collisions among a
fraction of the large-particle population may replenish the $\mu
m$-size IR-emitting particles.

The considerations presented in this subsection indicate that
the $\mu m$-size particle scenario suggested by Telesco {\it et
al.} (2000) and Wyatt {\it et al.} (2000) requires not only a
population of parent bodies which are larger than a few cm but
also a non continuous bimodal ($\sim 1 \mu m$ and $> 10 cm$)
particle-size distribution.  There is no simple physical mechanism
to preferentially produce this type of particle distribution
function.

\subsection{Particle size distribution}
In reality, collisions are more likely to produce fragments with a
range of sizes.  As noted above that all fragments with $a_d > a_1$
(in a gas-free environment) or $a_d > a_2$ (if the fragments are
embedded in a residual gaseous disk) would survive the radiative blow
out.  During the life span of HR4796A, the minimum-size parent
particles would undergo $\nu_{c I} \tau_\star \sim
(L_{disk}/L_\star)(\tau_\star/ \tau_d) \sim 10^3$ collisions.
Although fragmentary particles with $a_1 < a_d < a_{min}$ or $a_2 <
a_d < a_{min}$ may collide less frequently, they may nevertheless
attain a collisional coagulation-fragmentation equilibrium.

In principle, the particles' size distribution may be obtained by solving
the appropriate coagulation equation with both coagulation and fragmentation
effects included. Approximate solutions may be obtained for collisional
steady state in which the disruption or growth rates ($\dot M_d$) of the 
total mass ($M_d = N_d m_d$) for a given population may be assumed to be 
independent of $a_d$. If we assume the size-distribution evolution
occurs primarily through collisions between similar size particles,
\begin{equation}
\dot M_d \simeq M_d \nu_c \sim N_d ^2 m_d (a_d/R)^2 \Omega.
\end{equation}
A constant $\dot M_d$ would imply $N_d \propto a_d ^{-2.5}$, $M_d \propto
a_d ^{0.5}$ and $\tau_z \propto a_d^{-0.5}$, {\it i.e.} most of the mass
are contained in the large particles whereas most of the surface area cross
sections are contained in the small particles. This particle size
distribution is similar to that derived for collisional equilibrium
(Hellyer 1970, Mathis {\it et al.} 1977) and adopted by Augereau
{\it et al.} (1999) in which
\begin{equation}
d N_d/ da_d \propto a_d ^{-3.5}.
\label{eq:sized}
\end{equation}
More generally, the growth and fragmentation may be mainly regulated by
collisions between different-size particles, which would result in
a slightly modified size distribution.

The distribution function in eq(\ref{eq:sized}) is only applicable
for particles with $a_d \sim a_1$ or $a_d \sim a_2$.  Since smaller
particles are subject to radiative blow out of the ring before they can collide
with each other again, their size distribution is probably determined
by that resulting from the break up of the parent bodies.  But if
this distribution function can be extended to the small particles
the total mass
carried by the blown-away dust particles is $M_{lost} ^{tot} \sim
M_{lost} ^{char} (a_1 / a_c)^{1/2} \simeq 5-10 \times 10^{2} M_\oplus$ in the gas free
environment and $M_{lost} ^{tot} \sim M_{lost} ^{char} (a_2 / a_c)^{1/2}
\simeq 3 \times 10^{3} M_\oplus$ if the particles are embedded in a protostellar
disk of gas.  Note that the inferred values of $M_{lost} ^{char}$ and
$M_{lost} ^{tot}$ are comparable to Jupiter and Saturn's masses and
the Oort's cloud (Duncan, Quinn, \& Tremaine 1987). This mass is larger than 
the present mass estimate ($\sim 0.1 M_\oplus$) for the Kuiper Belt today 
(Luu {\it et al.} 1997) but may be comparable to it in the past (Stern {\it 
et al.} 1997, Kenyon \& Luu 1999).  The associated surface density of heavy 
elements is also comparable to that extrapolated from the minimum mass nebula 
for $R_0 = 70$AU.

\subsection{Preserved particle scenario}
We now consider the possibility that a range of particles were formed 
inside a gaseous protostellar disk.  When the surface density of the disk 
is depleted, particles in the outer regions of the disk become exposed to 
the stellar radiation, all particles with $a_d < a_2$ would be blown away 
by the radiation pressure.  If these small particles are not the dominant IR
emitters, they would not not need to be continually replenished by some
parent particle.  In their model analyses of a large wavelength range of 
IR emission, Augereau {\it et al.} (1999) adopt a distribution of particles 
with sufficiently large size to survive radiation blow outs.  We follow their
approach by examining the evolution of the surviving particles.

Using our model parameters, we find $a_2 = 100-200 \mu m$ (see Fig.\ \ref
{modA.ref} and Fig.\ \ref{modB.ref}).  (Note that $a_2$ does not directly 
depend on the gas surface density as long as $a_0 > a_2$). For illustrative 
simplicity, we assume the IR-emitting particles have a characteristic size 
$a_d = a_0 = 600 \mu m$ such that eq(\ref{mpart}) gives $M_{char} \sim 2.3 
M_\oplus$.  Since these particles survive the radiation blow out, they 
collide up to $\sim \nu_c \tau_\star \sim (L_{disk}/L_\star) (\tau_\star/
\tau_d) \sim 10^3$ times within the life span of the central star.  

In order to determine the collisional velocity among the particles, we note 
that in the absence of any warp, the optical depth in the disk's radial 
direction to the host star is $\tau_r \sim \tau_z R/H_d$.  Since the surface 
brightness of the disk is spread out over an extended radial range, $\tau_r 
< 1$ and $H_d / R > L_{disk}/L_\star$ for the IR emitting particles.  The 
corresponding velocity dispersion of the particles would be $\sigma_d
\geq 3 \times 10^3$cm s$^{-1}$.  If these particles are embedded in a 
gaseous background, $H_d$ is expected to be $\leq H_p$ which corresponds 
to an upper limit $ \sigma_d \leq 10^5$cm s$^{-1}$.  Within this velocity 
range, collisions may lead to either erosion (Hatzes {\it et al.}
1991) or coagulation if the particles are coated with frosts or traces
of methanol as in comets Supulver ({\it et al.} 1997).  Repeated
collisions may eventually lead to an equilibrium in which
fragmentation and coagulation are balanced to result in a size
distribution similar to that in eq(\ref{eq:sized}) (see review by 
Artymowicz 1997).

In addition to the mass requirement, a population of surviving
particles (with $a_d > a_1$ and $a_d > a_2$) need to have a spatial
distribution similar to that observed ring structure whether they are
parent particles or the dominant IR emitting particles.  In the following
subsection we show that particles which are marginally larger than
$a_2$ can congregate into rings if there are some non monotonic
variations in the $\Sigma$ distribution of the gas.  However, if
particles become larger than $a_0$, they decouple from the gas flow.
Nevertheless, this confinement mechanism continues to be effective
provided $dV$ become positive although the rate of concentration
decreases with the particle size.  Meter sized particles need $\approx 
10^3$ times longer to concentrate as they radially drift about $10^3$times 
slower then the $600 \mu$ particles. 

\subsection{A set of quasi-static models}

\subsubsection{Model: A\label{Sect_modA.ref}}
We first consider the hydrostatic structure of both gas and dust disk.
We set the initial surface density distribution of the gas to be
proportional to $R^{-5/2}$ as it is typical for protostellar
disks. The presence of gas is invoked in order to establish a flow
with velocity reversal which would spontaneously lead to the emergence
of ring structure.  In \S2.1, we adopt the surface density
distribution of the gas disk to be that in eq (1) with its mass within
$70 \pm 20$ AU to be $99 M_\oplus$ in order to augment for the solar
composition.  As a free parameter, we set $d=0.9$ which corresponds to
a smooth $10\%$ surface density increase at 70 AU.  The amplitude of
this variation in $\Sigma$ (see Fig.\ \ref{modA.ref} and Fig.\ \ref{modB.ref})
is much smaller than that associated with gap formation.

The disk structural parameters are calculated self-consistently for a
dynamically stable, rotating, optical-thin disk. This equilibrium
configuration is established on the dynamical time-scale $\tau_d$.  The
increase in surface density of the gas at 70 AU corresponds to a minor
peak in density, a local minimum in the friction time and stokes
number, but it has no effect on temperature and pressure scale height,
because the disk is optical thin (Fig.\ \ref{modA.ref}).  The pressure as function
of density also shows a small amplitude local maximum.  The resultant
local maximum in the radial pressure gradient strongly influences the
local rotational profile such that the deviation from the Keplerian
rotation speed is decreased by $50\%$.  The velocity resulting from
the Poynting Robertson effect is orders of magnitude smaller and it
attains this local minimum as a consequence of the decrease in the
friction time $\tau_f$ (see eq. 5).  The last frame shows the minimum
radius for particles to be gravitationally bound ($a_2$), which is
given in eq(\ref{eq:a_2}) by the equilibrium of the radiation
pressure, gravity and centrifugal forces. This plot also indicates 
the upper particle size limit for particles to be stopped at 
70AU, which is about $700-800 \mu$.

For these disk parameters, our chosen particle's size is ($a_d = a_0 = 600 \mu
m$) to receive maximal velocities and thus to show the upper limit of concentration.  
In Fig.\ \ref{modA_600.ref}, the solid line represents the
resulting drift velocity of these particles due to combined effect of
gas drag and radiation pressure in accordance to eq(\ref{Eq_v_r}).  In
order to consider the relative importance of various contributions, we
compare our main results with the dashed-line velocity curve for
particles in a gravity- and rotation-free system, where radial
radiation pressure and friction forces cancel each other.  In this
idealized case, the particles' radial velocity is positive everywhere
because the friction time grows faster than the radiation pressure
decreases with radius.  The decline in $v_r$ at 70 AU is due to the
local minimum in the friction time-scale.

We also show with the dotted line, the $v_r$ distribution in a system
where the radiation pressure is neglected and the orbital evolution is
determined by gas drag only. In this limit, the radial drift is
proportional to the friction time and the particles migrate inward
everywhere. Due to the non monotonic variation in our prescribed
$\Sigma$ distribution near 70 AU, the deviation of the gas velocity
from the Keplerian flow is at a minimum.  This change in $dV$ leads to
a large increase in $v_r$.  But with $d=0.9$ and $\delta R_0=20 AU$ 
in Model A, gas drag
alone is not sufficient to change the sign of $dV$ and $v_r$.
In contrast, when the effects of radiation pressure and gas drag are
fully taken into account simultaneously, a radial velocity inversion 
is established near 70 AU.

For this model, particles with $a_d < a_2$ are blown away by the
radiation pressure on the dynamical time-scale $\tau_d$.  When the
surface density of the gas is reduced below $\Sigma_2 \equiv 3
L_\star/ (8 \pi R c c_s^3 \partial {\rm ln} \rho / \partial {\rm ln} R)
\sim 0.1$g cm$^{-2}$, $a_0$ becomes smaller than $a_2$ 
(see eq.\ \ref{a_0} and eq.\ \ref{eq:a_2}) and there are no 
longer particles that drift radially outward as a result of the 
combined forces from gas and radiation.
If the motion
of the disk gas is only affected by a negative pressure gradient, all
particles with $a_d < a_0$ would be blown away by the stellar
radiation pressure and all particles with $a_d > a_0$ would become
segregated from the motion of the gas, drift inwards with a radial
velocity in accordance to eq(\ref{Eq_v_r>}) and a corresponding
orbital evolution time-scale $R/v_r \sim (R/H_p)^2 (a_d \rho_d /
\Sigma) \tau_d \sim (R/H_p)^2 \tau_d$. In this case, all particles
with sizes smaller than $\sim a_2 (H_p/R)^2 (\tau_\star/\tau_d)$
would be depleted during the lifespan of HR4796A.  But $\Sigma_2$ is
only marginally larger than the surface density of the parent
particles inferred in \S3.1.  In \S2.4, we have already shown that
the particles' drag have a negligible effect on the motion of the gas.
Eventually, when the surface density of the gas is reduce below
$\Sigma_1 \equiv 3 L_\star / (8 \pi G M_\star c) \sim 10^{-2}$g
cm$^{-2}$, $a_0 < a_1$.  All particles with $a_d < a_1$ are blown away
by the radiation pressure whereas larger particles would not undergo
further orbital decay as the effect of head wind diminishes.

\subsubsection{Model: B \label{Sect_modB.ref}}
We also computed a model with the same $d$ but a smaller $\delta R_0=10 AU$
which represents a gaseous ring with a 2 times steeper surface density
gradient (see Figs.\ \ref{modB.ref} and \ref{modB_600.ref}).  In this case, the gas rotation is hyper Keplerian with a
sign change in $dV$ such that gas drag alone may lead to an outward
velocity near 70 AU. All particles with $a_d > a_2$ or $a_d > a_0$
would migrate outwards in regions with a positive pressure gradient
and inwards in regions with a negative pressure gradient. In this case
there is no upper limit for the particle size in order to be stopped at
70 AU.
Fig.\ \ref{modB_1m.ref} shows that the resulting drift velocities for 1m
sized objects is about three orders of magnitude smaller which nevertheless
allows these bodies to concentrate in reasonable time in a ring at 70 AU.


\subsection{Dynamical evolution} 
The resulting radial drift velocity (Eq. \ref{Eq_v_r}) can be used
with the continuity equation to compute the global evolution of the
dust surface density $\Sigma_d$ as a result of their orbital migration.
In a cylindrical coordinate system,
\begin{equation} 
\frac{\partial \Sigma_d}{\partial t} = - \frac{1}{R} 
\frac{\partial}{\partial r} \left( R j_{\Sigma}\right).
\label{eq:evol}
\end{equation}
The local flux of dust particles is partially contributed by the
systematic radial drift $v_r$ (see Eq.\ \ref{Eq_v_r}).  Small
particles with $a_d < a_0$, are well coupled to the disk gas.  If the
disk gas is turbulent, as in accretion disks in many astrophysical
contexts (cf.\ Lin \& Papaloizou 1996), these particles may undergo
diffusion following the trace of turbulent eddies.  For computational
convenience, we describe the turbulent stirring as passive diffusion
in the conventional Navier-Stokes equation such that
\begin{equation} 
j_{\Sigma} = \Sigma_d v_r - \frac{D}{R} \frac{\partial}{\partial r} \left( R 
\Sigma_d\right).
\end{equation}

A similar approach was used by Morfill and V\"olk (1984) to study the 
reprocessing of meteorites in the solar nebula.  
Following Cuzzi {\it et al.} (1993), we assume the turbulent diffusivity
$D=\nu/S\!c$ with the Schmitt number which describes the coupling between 
the particle and the gas is set to be $S\!c = (1 + S\!t)^{1/2}$.
For the gas turbulent viscosity, we adopt the {\it ad hoc} 
$\alpha$ prescription which is commonly used in accretion disks (Shakura
\& Sunyaev 1973) such that
\begin{equation} 
\nu = \alpha c_s H_p.
\end{equation}
The Stokes number for particles is modified from that for a laminar disk
(see \S2.2.1) by 
\begin{equation} 
S\!t = \tau_f \omega_t.
\end{equation}
$ \omega_t$ is the circular frequency of the largest eddy
(typically smaller than
$\Omega$). Thus, from the definition of the effective viscosity it follows: 
\begin{equation} 
\omega_t = \frac{\nu}{H_p^2}.
\end{equation}
For the particle size we are considering, the turbulent Schmitt number
is $\sim 1$ so that maximum mixing is expected to occur.
 
The differential equation (\ref{eq:evol}) is solved numerically using
a finite volume scheme.  It is integrated explicitly in time with a
simple Euler backward scheme.  

\subsubsection{Model: A}
In Figure \ref{modA_dyn.ref} the initial radial
$\Sigma_d$ distribution of $600 \mu m$ dust particles ($= 10^{-2}
\Sigma$) is shown with the solid line.  The evolution of the
$\Sigma_d$ distribution is shown with a series of dotted lines
which represents time sequences separated by $10^3$ yr between each
line.  After the first $400$ yr, the dust particles inside 66AU drift
inward while that between 66 and 69 AU drift outward.  The lack of
replenishment causes $\Sigma_d$ interior to 66AU to decline.  The
results in Figure \ref{modA_600.ref} indicate that $v_r$ is a decreasing function of
$R$ (i.e. the magnitude of $v_r$ is more negative at large $R$) inside
66 AU.  Consequently, the depletion of the disk is more rapid at
larger R and $\Sigma_d$ becomes a decreasing function of $R$ in
this region.  The outward drift at 66-69AU and inward drift at radius
larger than 69AU cause the accumulation of dust particles between
67-73 AU and the depletion of dust exterior to 73 AU.  After several
$10^3$ yr, an equilibrium is reached in which the turbulent diffusion is
balanced by the systematic drifts.  In this equilibrium,
$\Sigma_d R$ between 67-73 AU becomes nearly two orders of magnitude
larger than that of the disk interior to 67 AU and another two orders
of magnitude larger than that of the disk beyond 73 AU.  The actual
width of the ring depends on the magnitude of the effective viscosity 
parameter.  A much wider ring may be obtained for a larger viscosity $\alpha > 10^{-4}$.

After the equilibrium is established, the persistence of the gas is
not required.  In \S2.2.2, we show that as $\Sigma$ decreases during
the depletion of the disk, $a_0$ decreases and $\tau_f$ increase while
$dV$ is not significantly modified.  While $a_d$ is still small
compared with $a_0$, the factor $(L_{\star} A_d/4 \pi R^2 m c + 2
\Omega dV)$ in eq(\ref{Eq_v_r}) remains unchanged.  Although the
magnitude of $v_r$ may increase due to the weakening hydrodynamic drag
effect, the locations for velocity reversal remain unaltered.  Thus,
dust particle concentration in the ring is not affected by the
depletion of $\Sigma$.  Although dust particles would generally
migrate inwards after $a_0$ decreases below their size $a_d$, the drag
time-scale becomes so large that the radial locations of the particles
are preserved.  Thus, this scenario only needs the gas to be present
for the initial few $10^4$ yrs in order to produce the
ring. Afterwards it can be evaporated or be viscously diffused, while
the dust distribution remains intact.

\subsubsection{Model: B}
Similar model parameter are applied to a narrower ring (
with ${\delta R_0=10 AU}$).  In this model, we examine the 
concentration of 1$m$-size large objects.
Fig.\ \ref{modB_dyn.ref} shows the same behavior like the smaller
particles, but this time the process is about 1000 times slower.
Nevertheless, the concentration process takes an order of magnitude 
less time than the life time of HR4796A.

\subsubsection{Model: C and D}
So far the upper limit for the observed gas around HR4796A is
1-7 $M_\oplus$ (Greaves {\it et al.} 1999), which is 1-2 orders
of magnitude less than we used for Model A and Model B.
Now we present to models with only 10 $M_\oplus$ of gas (see fig.\ \ref{modC.ref})
and 1 $M_\oplus$ (see fig.\ \ref{modD.ref}). In comparison to the first models
the densities are smaller, the Stokes numbers for 600 $\mu$ particles are 
bigger and thus the resulting drift velocities become also smaller. 
There are no significant changes in the critical particle size $a_2$. 

Thus the dynamical evolution takes 5 times longer for Model C (see fig.\ \ref{modC_dyn.ref})
and about 25 times longer for Model D (see fig.\ \ref{modD_dyn.ref}).
Even the time-scales increase at lower gas densities, the
general effect persists. 

\section{Summary and Discussions}
We examined in this paper the dynamics of the dust ring around HR 4796A
with two separate scenarios.  In the first scenario, we consider the
case where the dominant emitters of the IR radiation are $\mu m$-size
particles.  Since these particles are being blown away by the
radiation pressure of the central star, they must be replenished by
large particles which are not affected by such a process.  In a
gas-free case, the parent particles must have a size greater than
$\sim$ 3.2 $\mu m$ whereas in a gas-rich environment, another order of
magnitude larger size is needed.  In order to account for the
magnitude of reprocess radiation, $M_{char} \sim 4-8 \times 10^{-3}
M_\oplus$ of $\mu m$-size particles are needed at any given time.
During the life span of HR4796A, a total mass $M_{lost}^{char} \sim
M_{char} \tau_\star /\tau_d \sim 300 M_\oplus$ is needed in the
reservoir of parent particles.  In order for the parent particles to
be inconspicuous sources of IR radiation relative to the $\mu m$-size
particles, a large fraction of $M_{lost}$ must be contained in
particles with $a_d > 10$cm.  However, the smallest surviving
particles (with sizes $\sim 100 - 200 \mu m$), could provide adequate
supplies of $\mu m$ size particles if a large fraction of their mutual
collisions results in their total fragmentation.

The main difficulty of identifying small particles as the dominant
contributors of the IR radiation is the requirement of a non continuous
bimodal size distribution.  This problem would be removed
if the particles which emit most of the IR radiation are sufficiently
large ($> 100 \mu m$) to survive the radiative blow out. In this case, a total
mass $< 0.4 M_\oplus$ is needed throughout the life span of HR 4796A.
We suggest the preservation of these modest-size particles (whether 
they are parent particles or the dominant IR emitters) is due to the
combined effect of radiation pressure and gas drag.

We presented a detailed model for the formation of dust rings in
circumstellar disks. Our model can account for rings with sharply defined edges
if it is embedded in a gas disk. A very small amount of inhomogeneity is needed
to induce the congregation of the particles.  This inhomogeneity 
can be the result of a planet, but it does not necessarily has to be
so.  Photo-evaporation, an inhomogeneous initial distribution of mass
after the infall, or inhomogeneities in the radial viscous transport
of material can also account for such an anomaly in the surface
density distribution.  Thus we conclude that dust rings with sharp 
structures are not necessarily good indicator for embedded planets. 

We thank D.\ Trilling and P.\ Bodenheimer for useful conversation. This work is supported
in part through National Science Foundation grant AST-9618548, NASA
grants NAG5-4277, 4494, 7515, and a special NASA astrophysics theory
programme which supports a joint Center for Star Formation Studies at
NASA-Ames Research Center, UC Berkeley, and UC Santa Cruz.

\clearpage

\figcaption[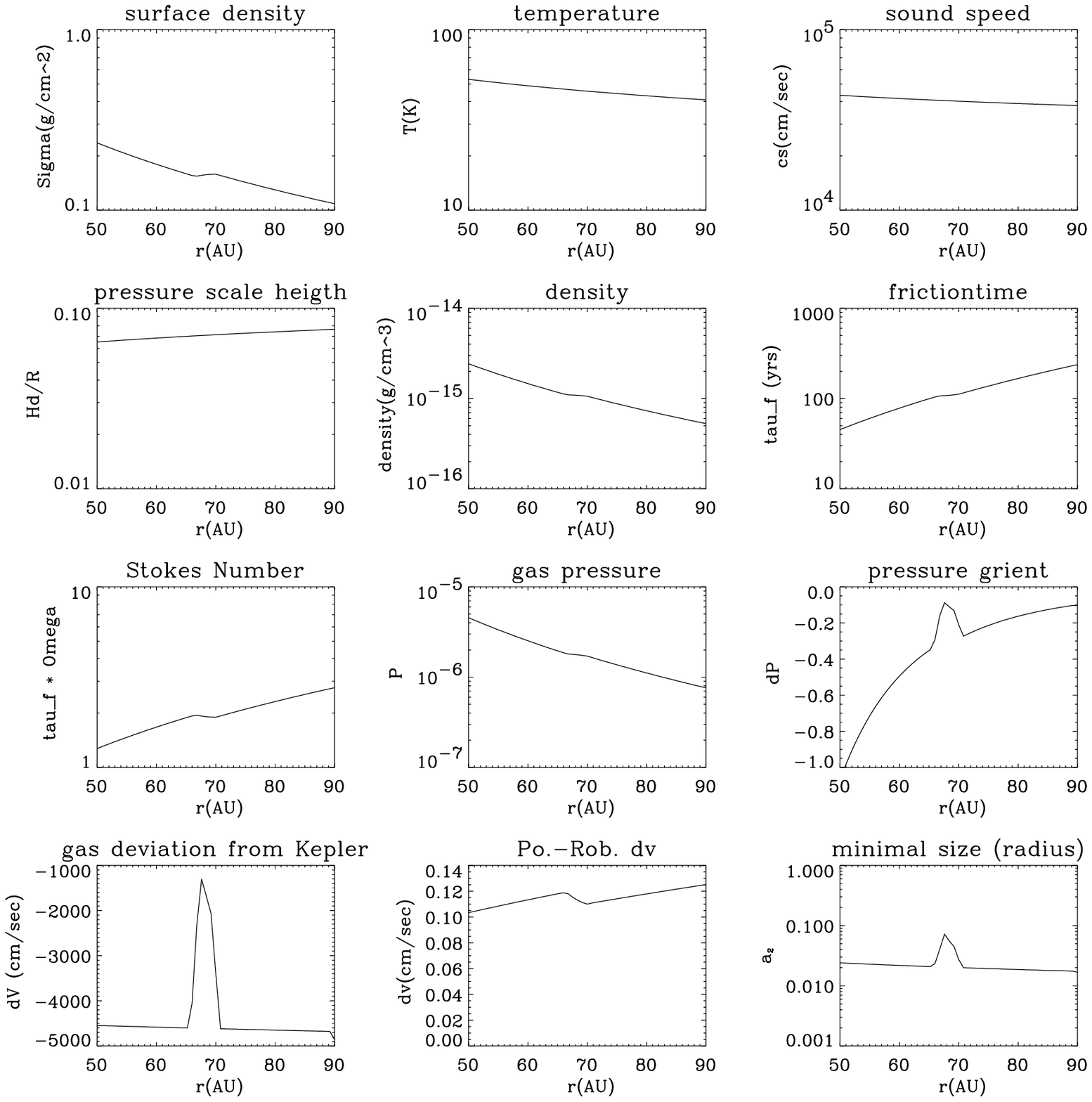]{\label{modA.ref}Model A: The radial distribution of
surface density, density, temperature, friction time for 600 $\mu$-size 
particles, the resulting Stoke-Number, pressure scale height, the speed 
of sound, the pressure, the radial pressure gradient, the deviation from 
the Kepler rotational profile, the Poynting-Robertson effect and the size (cm)
of the smallest gravitationally bound particles. The physical parameters for
this plot are given by the mass (${M_\star = 2.5 M_{\sun}} $) and luminosity 
(${L_\star = 35 L_{\sun}}$) of the central object. Additional model
parameters include the location of the maximum in surface density at 
$70 AU$, the width of the inner edge as ${\delta R_0=20 AU}$ and the gas 
mass around the edge between ${R_0- \delta R_0}$ and ${R_0+ \delta R_0}$ 
which is taken to be 100 earth masses.}

\figcaption[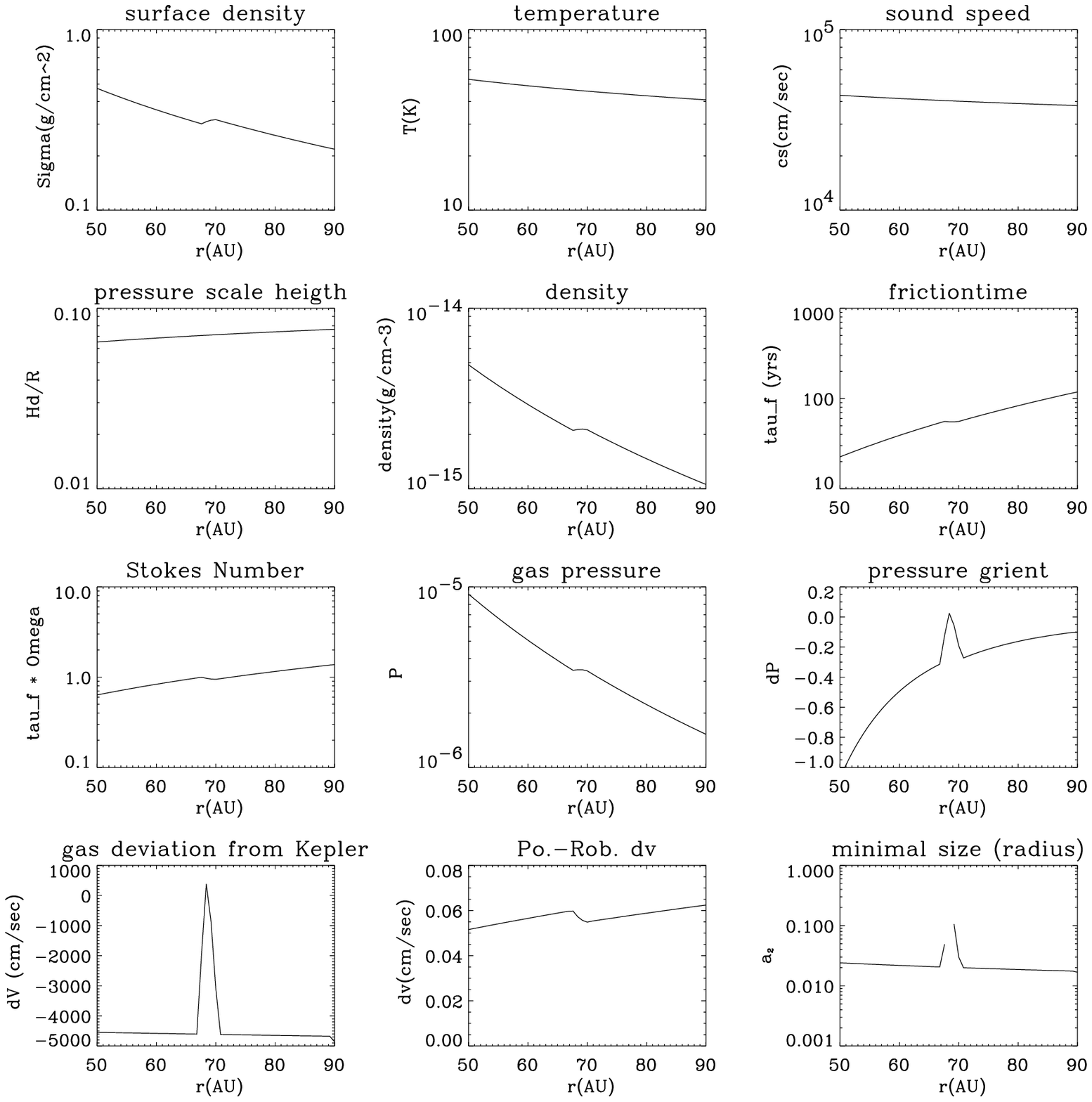]{\label{modB.ref}Model B: The radial distribution of
surface density, density, temperature, friction time for 600 $\mu$-size 
particles, the resulting Stoke-Number, pressure scale height, the speed 
of sound, the pressure, the radial pressure gradient, the deviation from 
the Kepler rotational profile, the Poynting-Robertson effect and the size 
of the smallest gravitationally bound particles. The physical parameters for
this plot are given by the mass (${M_\star = 2.5 M_{\sun}} $) and luminosity 
(${L_\star = 35 L_{\sun}}$) of the central object. Additional model parameters 
include the location of the maximum in surface density at $70 AU$, the width 
of the inner edge which is ${\delta R_0=10 AU}$ and the gas mass around the 
edge between ${R_0- \delta R_0}$ and ${R_0+ \delta R_0}$ which is 100 earth 
masses.}

\figcaption[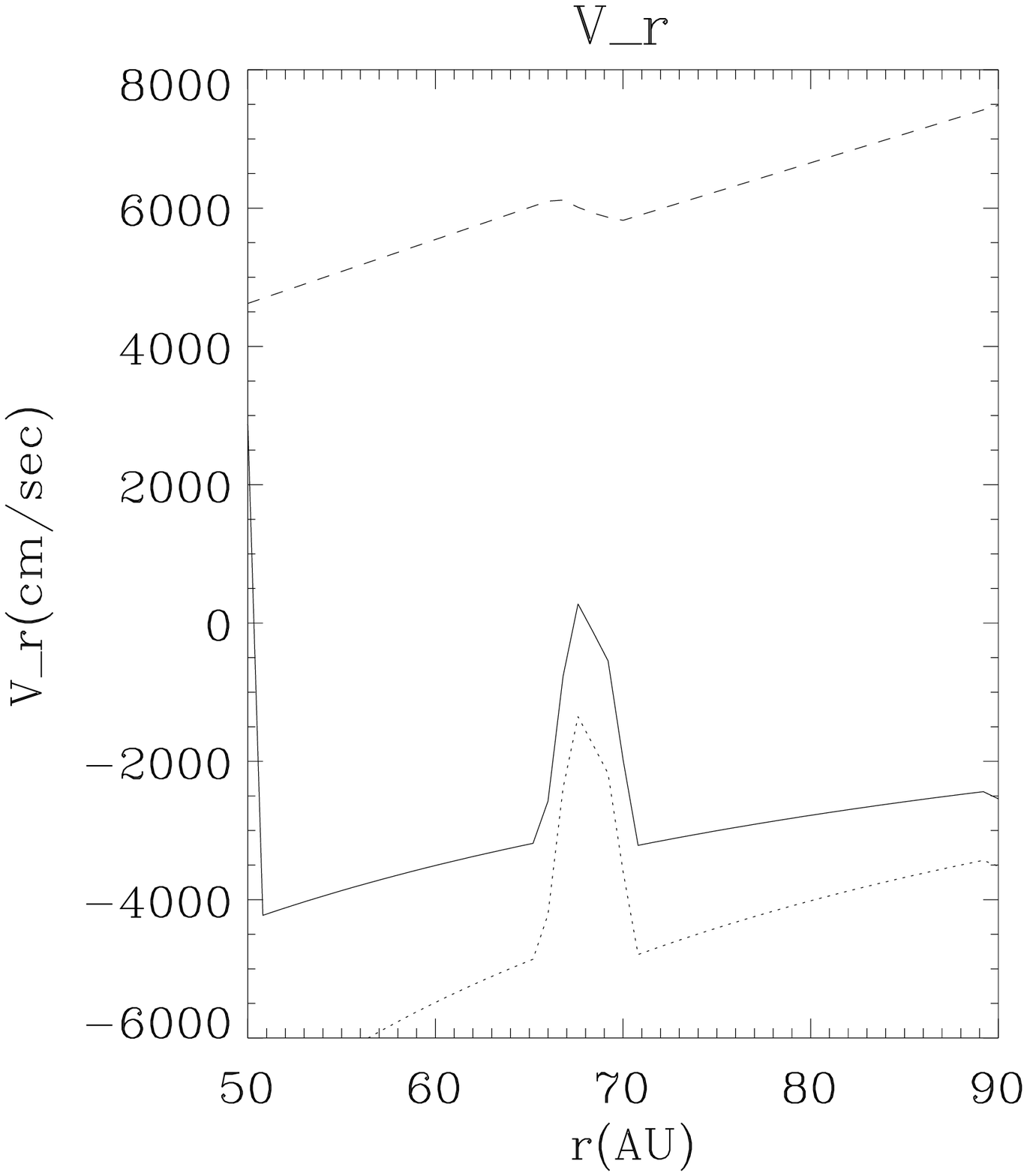]{Radial velocities distribution of 600 $\mu$-size 
particles for the model in Fig.\ \ref{modA.ref}.
The solid line gives the effective radial drift. The dotted line accounts
only for the gas-drag effect and the dashed line denotes only for the radiation
pressure effect. \label{modA_600.ref}}

\figcaption[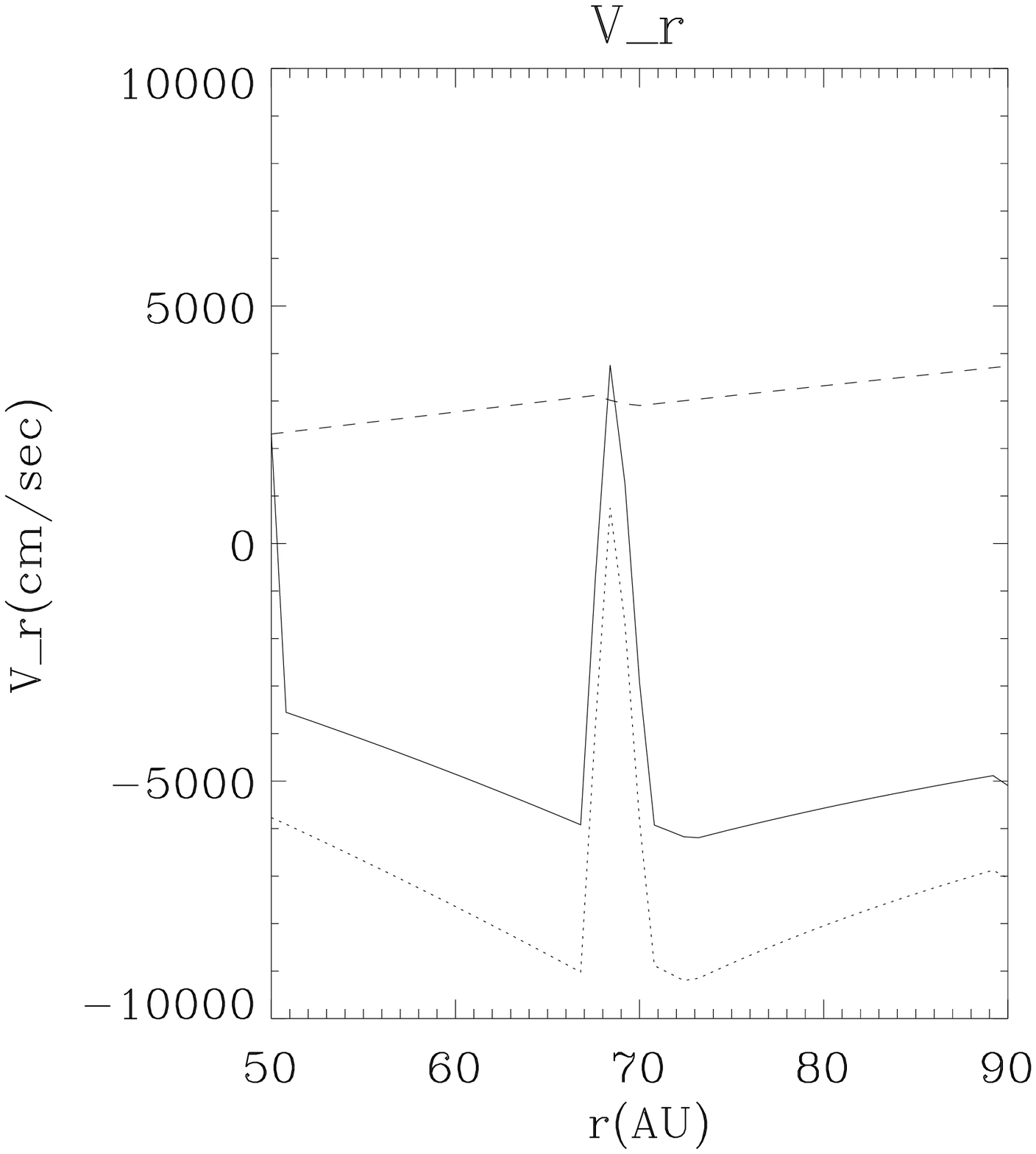]{Radial velocity distribution of 600 $\mu$-size 
particles for the model in Fig.\ \ref{modB.ref}.
The solid line gives the effective radial drift.\label{modB_600.ref}}

\figcaption[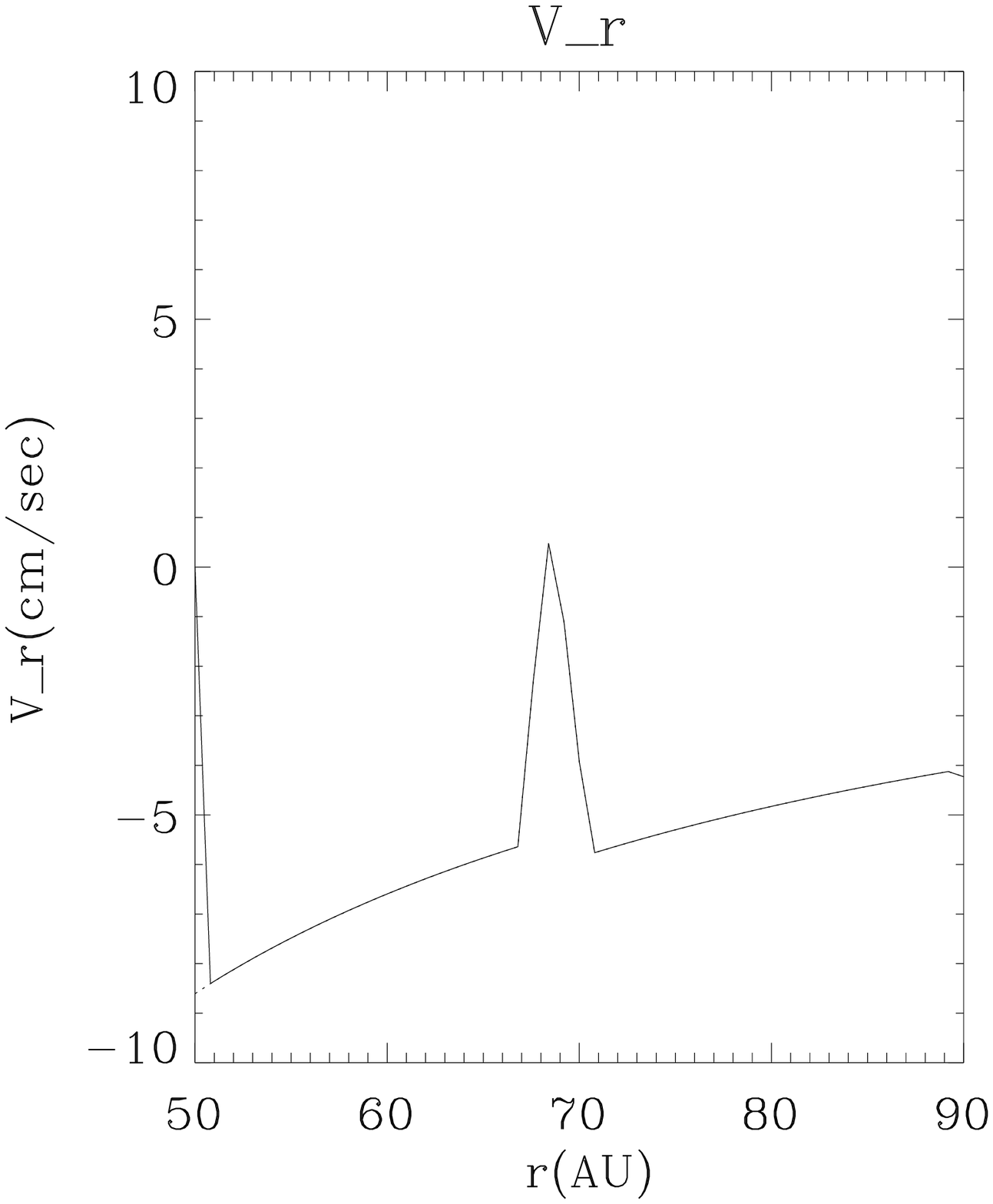]{Radial velocity distribution of 1 $m$-size 
particles for the model in Fig.\ \ref{modB.ref}.
The solid line gives the effective radial drift.\label{modB_1m.ref}}

\figcaption[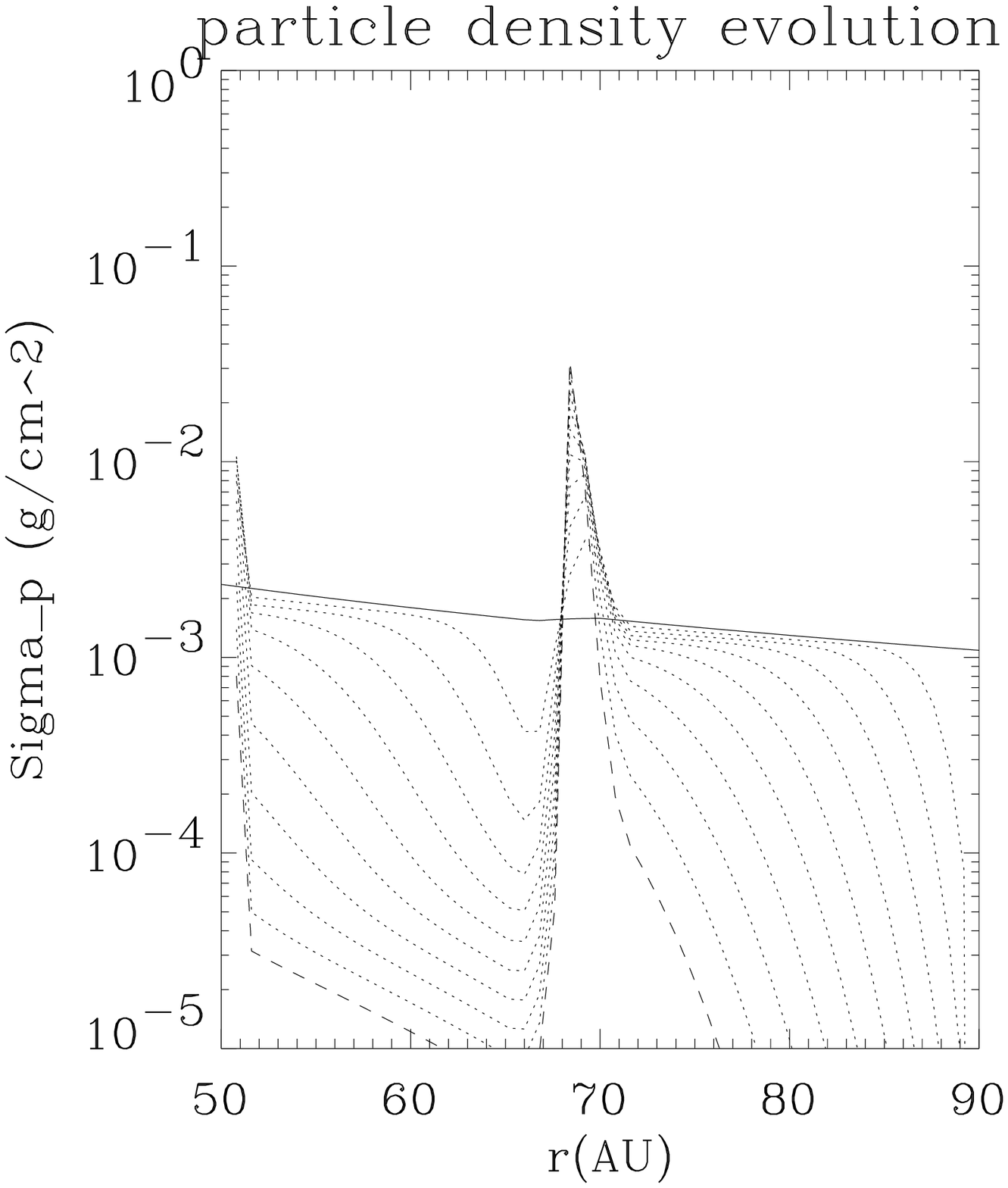]{\label{modA_dyn.ref}Model A:Evolution of the 
$600\mu$-size particle surface density distribution for the model in 
Fig.\ \ref{modA.ref}. The solid line gives the initial distribution. 
The following lines are snapshots every 400 yrs.}

\figcaption[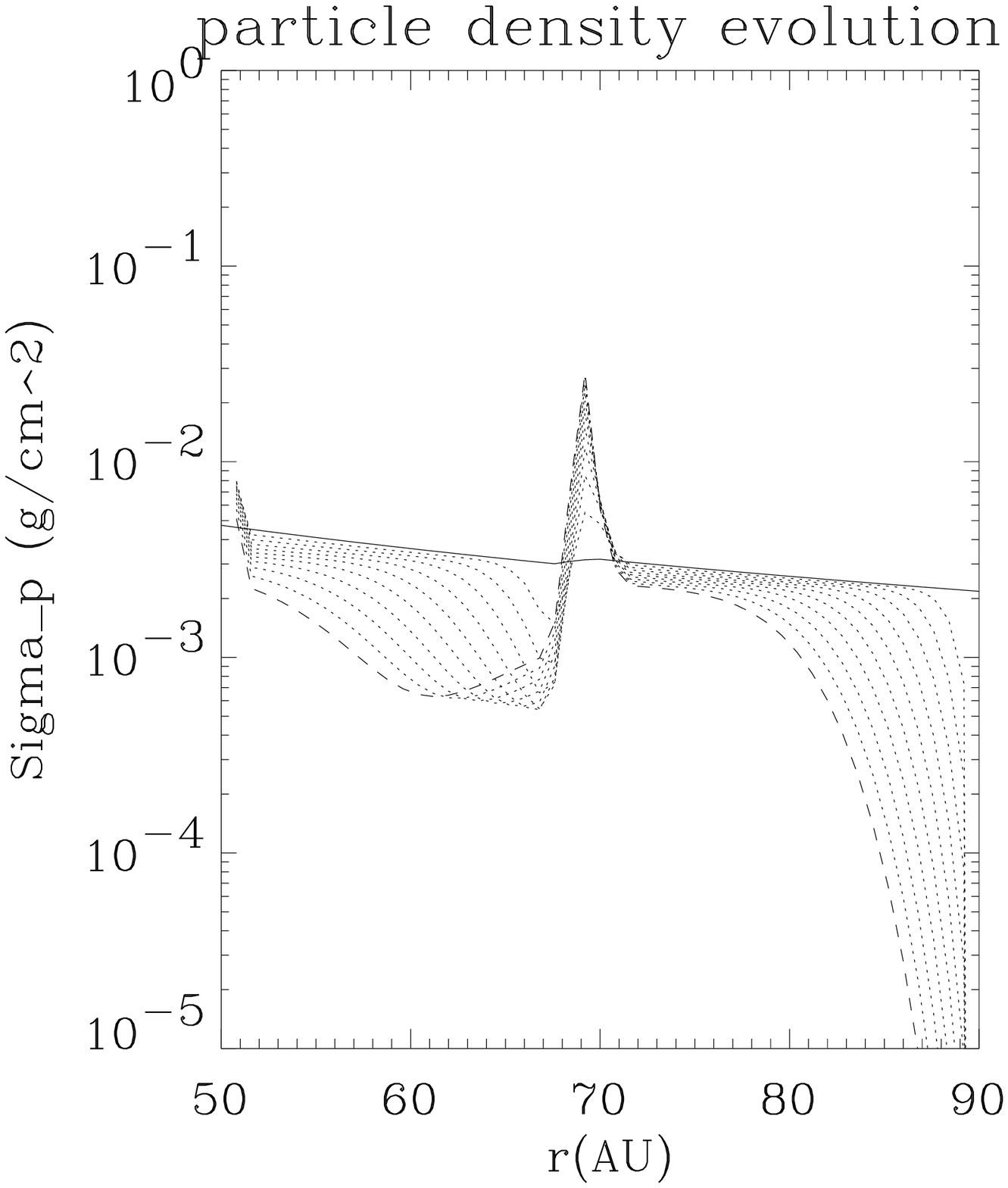]{\label{modB_dyn.ref}Model B:Evolution of the 
$1 m$-size particle surface density distribution for the model in Fig.\ 
\ref{modB.ref}. The solid line gives the initial distribution. The 
following lines are snapshots every $10^5$ yrs.}

\figcaption[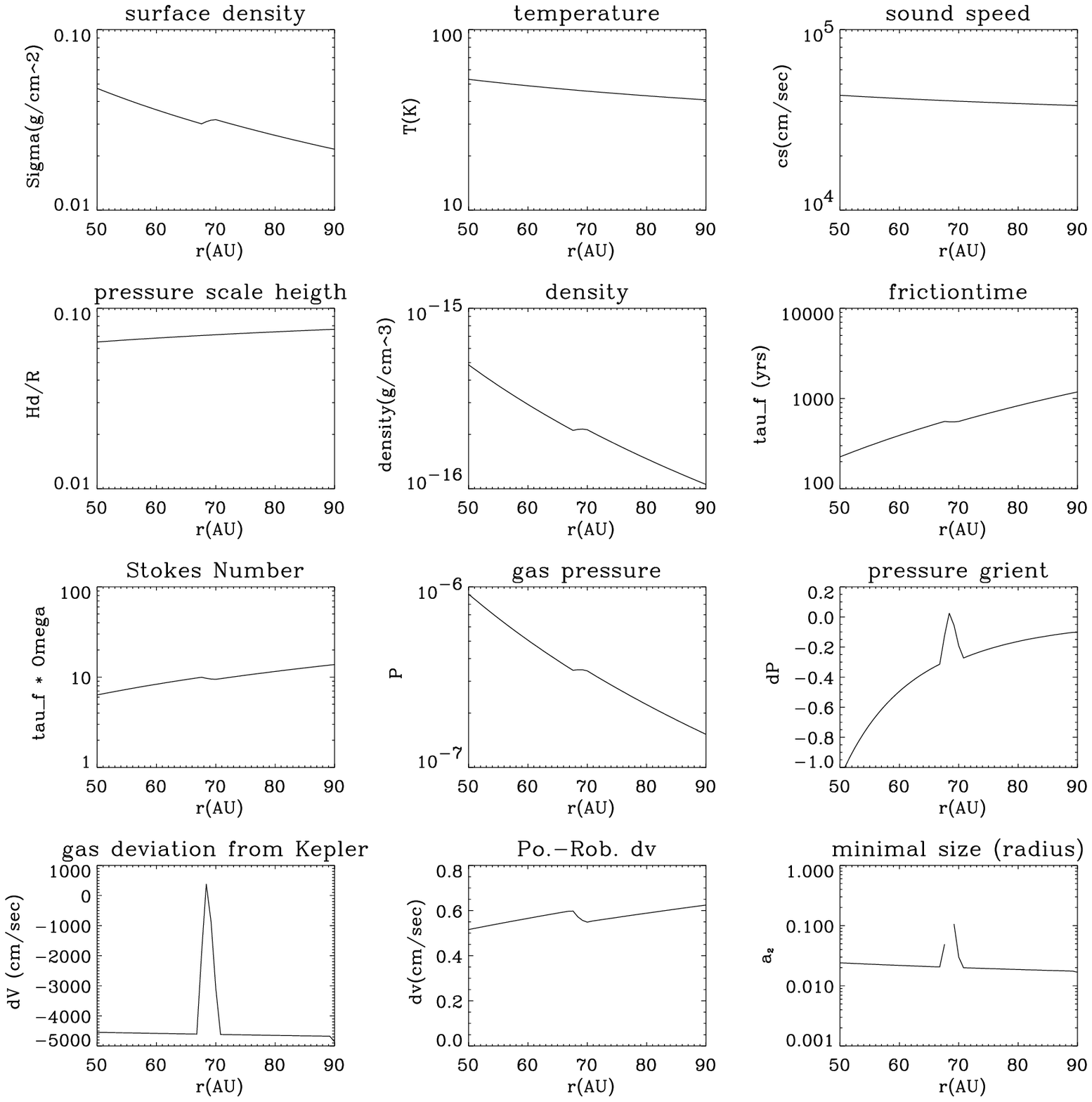]{\label{modC.ref}Model C: The radial distribution of 
surface density, density, temperature, friction time for 600 $\mu$-size
particles, the resulting Stoke-Number, pressure scale height, the speed 
of sound, the pressure, the radial pressure gradient, the deviation from 
the Kepler rotational profile, the Poynting-Robertson effect and the size 
of the smallest gravitationally bound particles. The physical parameters for
this plot are given by the mass (${M_\star = 2.5 M_{\sun}} $) and luminosity 
(${L_\star = 35 L_{\sun}}$) of the central object. Additional model parameters 
include the location of the maximum in surface density at $70 AU$, the width 
of the inner edge which is ${\delta R_0=10 AU}$ and the gas mass around the 
edge between ${R_0- \delta R_0}$ and ${R_0+ \delta R_0}$ which is 10 earth 
masses.}

\figcaption[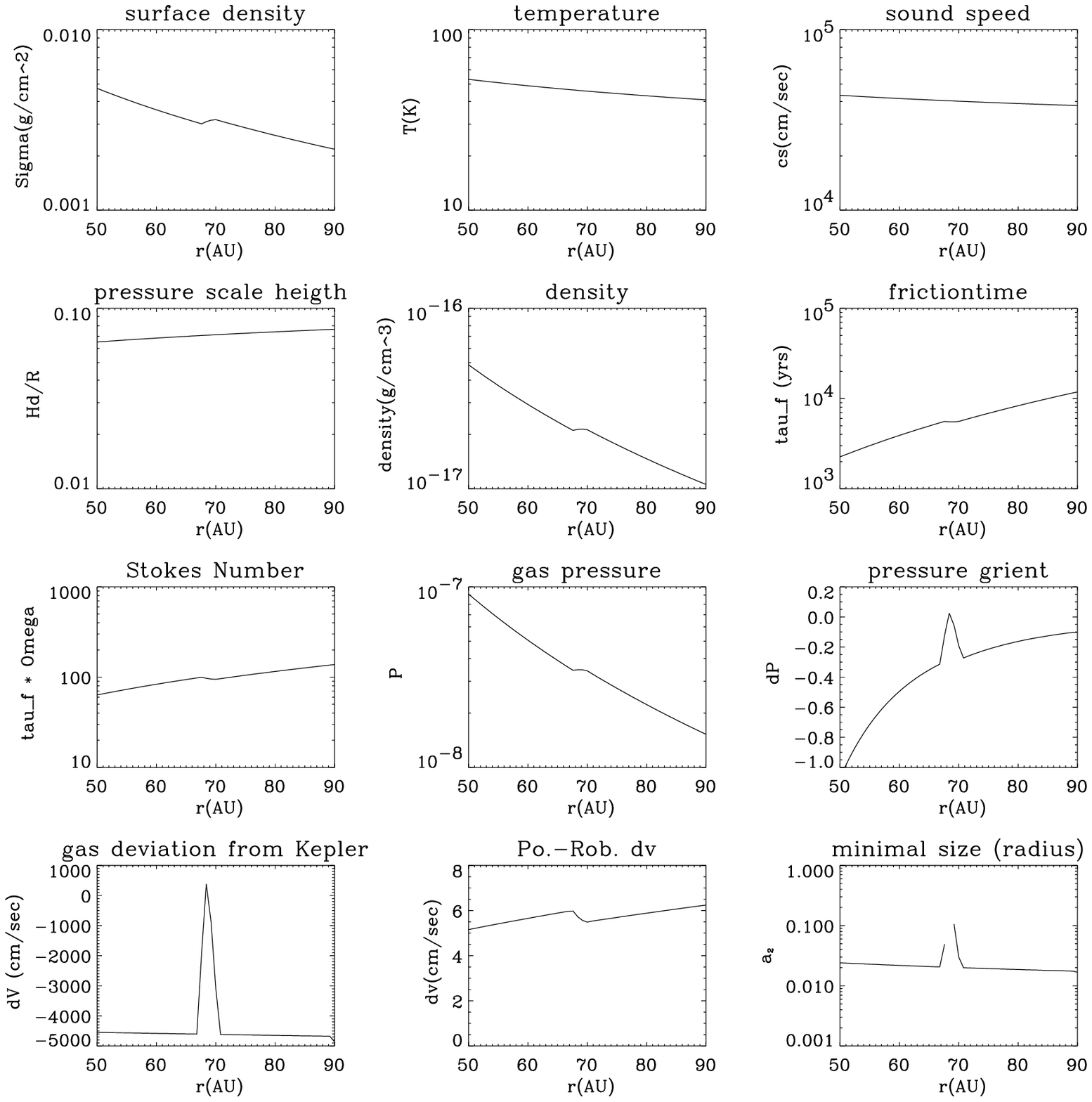]{\label{modD.ref}Model D: The radial distribution for 
surface density, density, temperature, friction time for 600 $\mu$-size 
particles, the resulting Stoke-Number, pressure scale height, the speed 
of sound, the pressure, the radial pressure gradient, the deviation from 
the Kepler rotational profile, the Poynting-Robertson effect and the size 
of the smallest gravitationally bound particles. The physical parameters for
this plot are given by the mass (${M_\star = 2.5 M_{\sun}} $) and luminosity 
(${L_\star = 35 L_{\sun}}$) of the central object. Additional model parameters 
include the location of the maximum in surface density at $70 AU$, the width 
of the inner edge which is ${\delta R_0=10 AU}$ and the gas mass around the 
edge between ${R_0- \delta R_0}$ and ${R_0+ \delta R_0}$ which is 1 earth 
mass.}

\figcaption[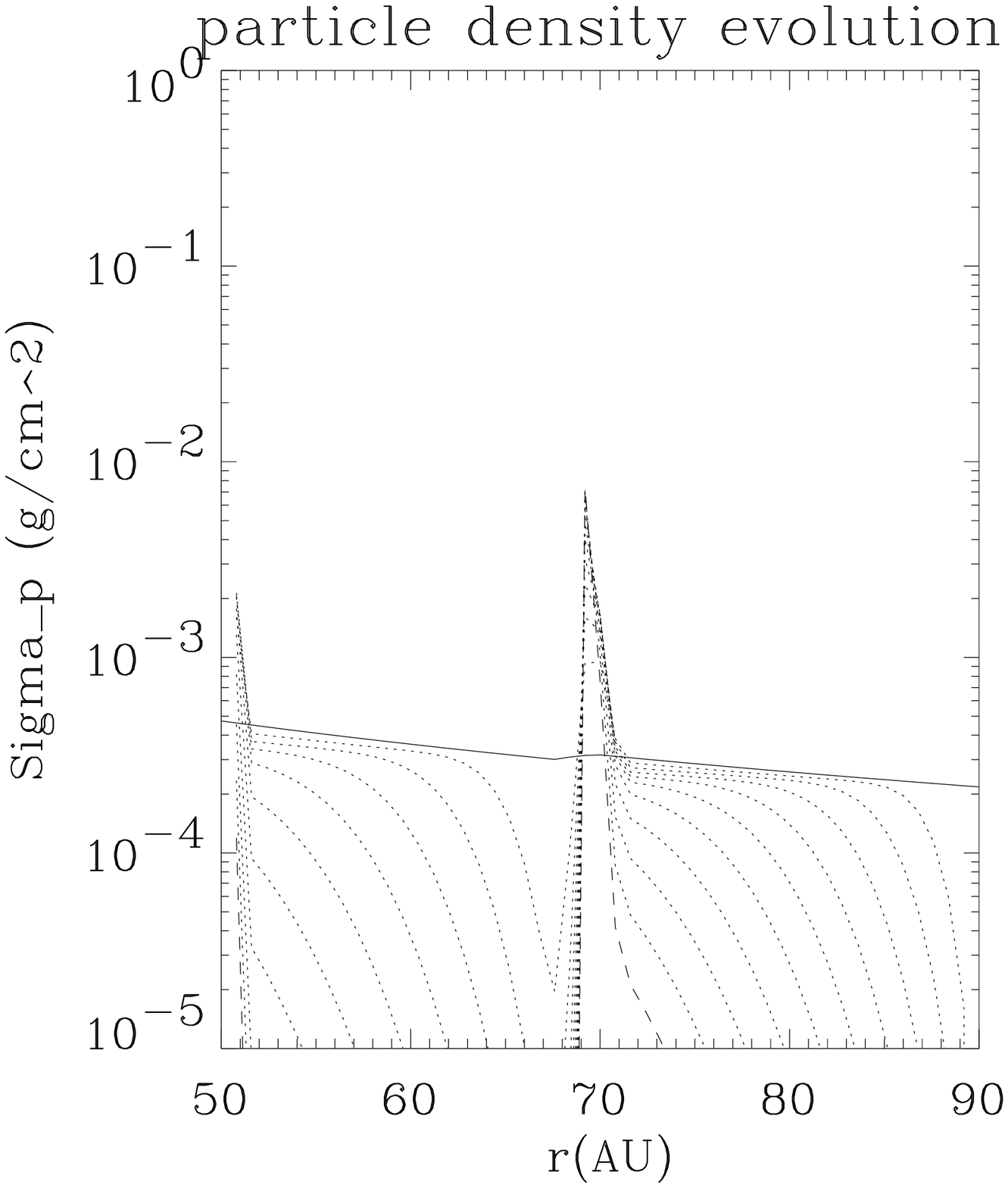]{\label{modC_dyn.ref}Model C: Evolution of the 
$600\mu$-size particle surface density distribution 
for the model in Fig.\ \ref{modC.ref}.
The solid line gives the initial distribution. Other lines are separated by
a time interval of 2000 yrs.}

\figcaption[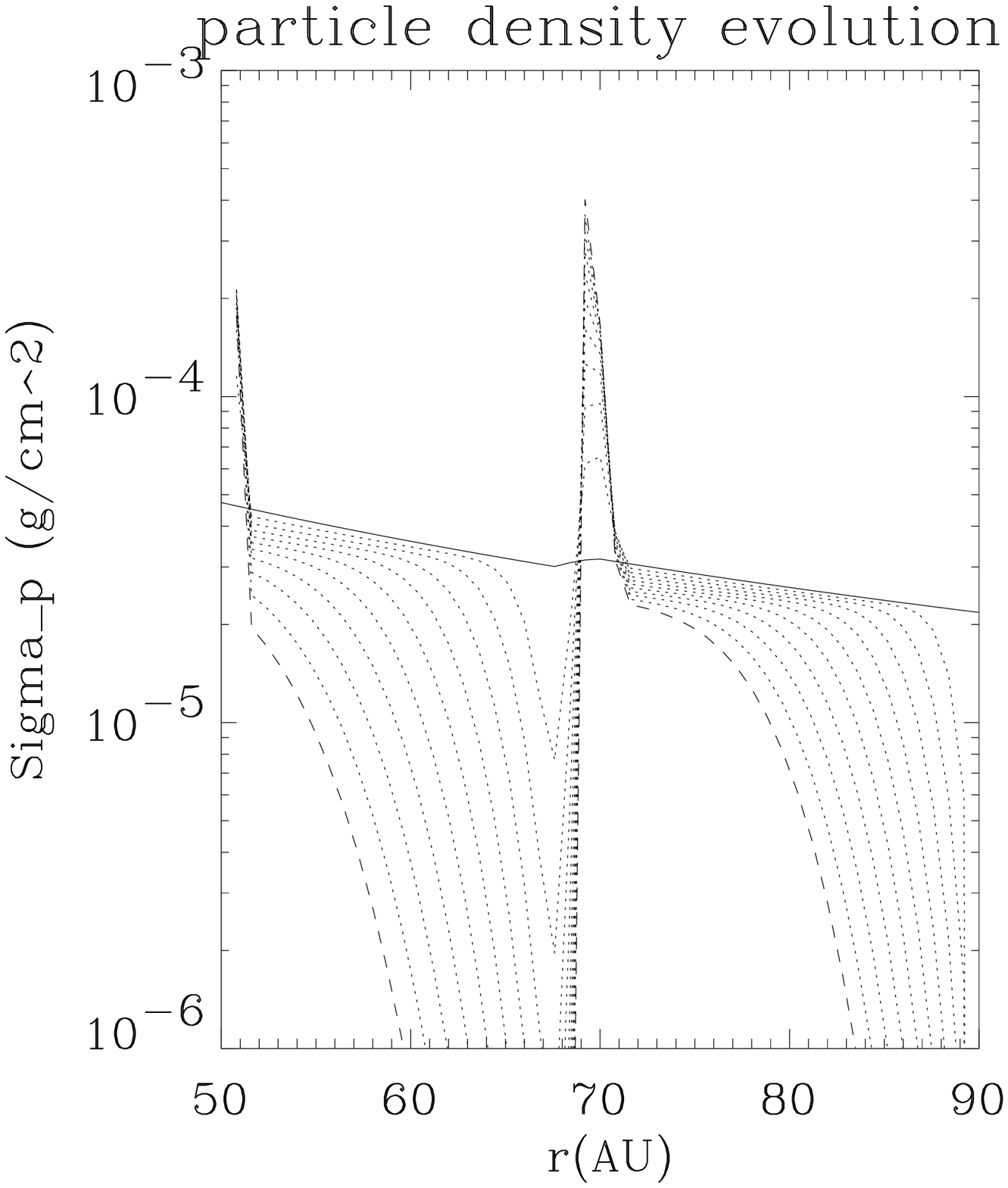]{\label{modD_dyn.ref}Model D: Evolution of the 
$600\mu$-particle surface density distribution
for the model in Fig.\ \ref{modD.ref}.
The solid line gives the initial distribution. The other lines are
separated by time intervals of every $10^4$ yrs.}

\clearpage
Fig.\ 1
\plotone{modA.ps}

\clearpage
Fig.\ 2
\plotone{modB.ps}

\clearpage
Fig.\ 3
\plotone{modA_600.ps}

\clearpage
Fig.\ 4
\plotone{modB_600.ps}

\clearpage
Fig.\ 5
\plotone{modB_1m.ps}

\clearpage
Fig.\ 6
\plotone{modA_dyn.ps}

\clearpage
Fig.\ 7
\plotone{modB_dyn.ps}

\clearpage
Fig.\ 8
\plotone{modC.ps}

\clearpage
Fig.\ 9
\plotone{modD.ps}

\clearpage
Fig.\ 10
\plotone{modC_dyn.ps}

\clearpage
Fig.\ 11
\plotone{modD_dyn.ps}
     
\end{document}